\documentclass[12pt, a4paper]{amsart}    

\usepackage[margin=0.9in]{geometry}
\usepackage[utf8]{inputenc}
\usepackage{amsmath, amssymb, amsthm, mathrsfs}
\usepackage[colorlinks=true, allcolors=blue]{hyperref}
\usepackage[foot]{amsaddr}
\makeatletter
\def\@setdate{Date: \@date\@addpunct.}
\makeatother

\def\equationautorefname#1#2\null{(#2\null)}

\usepackage[sort&compress]{natbib}     

\usepackage{graphicx}
\usepackage{pgfplots}
\pgfplotsset{compat=newest}
\usepackage{subcaption}
\usepackage{algorithm}
\usepackage{algorithmic}
\usepackage{booktabs}
\usepackage{multirow}
\usepackage{enumitem}

\usepackage[font=small,margin=0.5in]{caption}

\usepackage{setspace}

\DeclareMathOperator*{\softmax}{softmax}

\hypersetup{
    pdftitle={Deep Learning Across Games},
    pdfauthor={Daniele Condorelli and Massimiliano Furlan},
    pdfkeywords={learning in games, bimatrix games, neural networks, equilibrium selection},    
}

\title{Deep Learning Across Games}
\author{Daniele Condorelli and Massimiliano Furlan}
\thanks{\setstretch{1.0}\today, Department of Economics, University of Warwick --- Daniele Condorelli: d.condorelli@warwick.ac.uk; Massimiliano Furlan: massimiliano.furlan@warwick.ac.uk. We thank Shiva Chelluri and Maddy Pedder for research assistance and Joe Basford, Pierpaolo Battigalli, Kim-Sau Chung, Eduardo Faingold, Fran\c{c}oise Forges, Alkis Georgiadis-Harris, Kevin Leyton-Brown, Marco Li Calzi, Francesco Nava, Ludovic Renou, Balazs Szentes and Bernhard von Stengel for comments. We are additionally grateful to Bernhard for sharing his code to perform the Harsanyi-Selten linear tracing procedure. We also thank seminar audiences at the University of Warwick, HKU, HKBU, University of Bergamo, NYU Abu Dhabi, Bonn, University of Milan Bicocca, FGV (Rio and Sao Paulo), University of Chicago, University of British Columbia and at the Royal Holloway Theory Conference. The code is available at \href{https://github.com/massimilianofurlan/nn_bimatrix_games}{https://github.com/massimilianofurlan/nn\_bimatrix\_games}.}

\begin{document}

\begin{abstract}   

We train two neural networks adversarially to play static games. At each iteration, a row and column network observe a new random bimatrix game and output individual mixed strategies. The parameters of each network are independently updated via stochastic gradient descent on a loss defined by the individual squared regret experienced in the game. Simulations show the joint behavior of the trained networks approximates a Nash equilibrium in all games. In $2\times2$ games with multiple equilibria, the networks select the risk dominant equilibrium. These findings, which are robust and generalise out-of-distribution, illustrate how equilibrium emerges from learning across heterogeneous games.

\end{abstract}

\maketitle
     
\newpage


\section{Introduction}\label{sec:intro}                    

Nash equilibrium is the most widely used solution concept in applications of game theory. Yet concerns remain about its foundations. Even if players are mutually certain of their ability to best respond, what ensures correct beliefs are formed about their opponents'~behaviour? The textbook approach to resolve this issue is to interpret Nash play as the long-run outcome of a learning process or evolutionary dynamics.         

However canonical, this explanation is not fully satisfactory. Convergence may require many periods, whereas individuals never engage in repeated play of the same exact game (see Heraclitus {\textcolor{blue}{DK22B12}}\nocite{DK22B12} for an earlier conceptualisation of this observation). Moreover, equilibrium retains predictive power even when subjects encounter a game for the first time (see \citet{Goeree2001}). Therefore, it is common to informally hypothesise that humans learn by drawing analogies from different games. As \citet{FudenbergLevine98} note, ``our presumption that players do extrapolate across games [...] is an important reason to think that learning models have some relevance in real-world situations.''\footnote{Similarly, \citet{Kreps1990} writes: ``If we rely solely on the story of directly relevant experience to justify attention to Nash equilibria, and if by directly relevant experience we mean only experience with precisely the same game in precisely the same situation, then this story will take us very little distance outside the laboratory. It seems unlikely that a given situation will recur precisely in all its aspects, and so this story should be thought of as applying to a sequence of ‘similar’ situations.''}

In this paper, we propose a model of adversarial learning \textit{across} games and provide evidence that supports Nash play as a limit result, without postulating an exogenous notion of similarity among games or making agents play any one game repeatedly.\footnote{As usual, the assumption implying players do not attempt to influence the opponent's behaviour is motivated by the hypothesis that learning takes place within a large population of randomly matched agents.} Computationally, we show that the behaviour of two independent neural networks, trained with the goal of maximising their own expected stage-payoff while playing a sequence of observed randomly generated static games, converges towards a specific selection from the Nash correspondence.     

The model is as follows. A neural network row (or column) player is a function parameterised by a high-dimensional vector of weights, which maps each two-player static game into a mixed strategy for the row (or column) player.\footnote{The functional form is differentiable and flexible enough to approximate any function arbitrarily closely given a sufficiently large parameter vector  (e.g., see \citet{Hornik1989}).}
A row and column network are trained through a process that updates their randomly initialised weights across iterations. In each period a bimatrix game is sampled and fed to both networks, who then output a mixed strategy to play it. If a network does not best respond to the opponent's strategy, its weights are adjusted in a direction that locally increases the payoff in that game given the opponent's choice. The size of each period's move is regulated by a decaying learning rate.     

Because both networks see the bimatrix, the dynamic is not uncoupled in the sense of \citet{Hart2003}. Hence, our contribution is not to exhibit some universal Nash solver, such as the Lemke-Howson, but to show that a distributed, strictly adversarial meta-learning dynamic can converge to compute an approximate equilibrium in all games without playing any one game twice. The coupling we introduce is minimal, as adversarial as it can be with complete information: opponent's payoffs do not affect the sign of the ascent direction in behaviour space (i.e., the policy gradient). Moreover, if the stage game were fixed, then the process could be uncoupled for free. Finally,  without any signal about the opponent's payoff, convergence to Nash is provably impossible (see \citet{Foster2001}).

Two vectors of parameters that make the networks select a Nash equilibrium in every game represent a stationary point of our process. However, other local minima may exist. Despite payoffs being linear in the actions taken, the output of neural networks is not concave in parameters. In addition, there is no guarantee that playing Nash in every game is a global attractor of the dynamical system, which prevents us from using off-the-shelf stochastic approximation theorems. Indeed, to our knowledge, there are no theoretical guarantees that the iterates converge to prescribing Nash play in the environment we consider  (or in a closely related one). These considerations motivate our numerical approach.

For a variety of random seeds and architectural scenarios, we performed the computational training of paired neural networks on one billion of \(2 \times 2\) and,  separately, \(3 \times 3\) games. Then, we evaluated the performance of the last-iterates on different sets of \(2^{17}\) (\(\approx\) 130,000) randomly generated games. 
Individual payoffs are generated using float32 numbers with mean zero and unit variance.\footnote{For perspective, there are around 131 billion distinct \(3\times 3\) games considering strict \textit{ordinal} preferences over outcomes. A rough calculation suggests a single randomly generated float32 number is able to take on millions of possible values. Hence, the probability of encountering during training two games where payoffs represent the same exact preferences for any of the players is virtually zero, both in \(2\times 2\) and \(3\times 3\) games.} The average difference between the highest and lowest individual payoff in a game is 2.58 in \(2 \times 2\) games and 3.25 in \(3 \times 3\).

We find that the networks learn to approximate Nash equilibrium in nearly every game, with precision increasing along the learning curve. 
An \(\epsilon\)-Nash equilibrium is achieved if the maximum regret experienced by either player normalised by the difference between the max and min payoff in the game (MaxReg) is smaller than \(\epsilon\).  In our baseline simulations, after training on one billion \(2 \times 2\) games the average MaxReg in the test set is \textit{nearly zero}. The average MaxReg is less than \(0.01\) in \(3 \times 3\) games, after training on a billion games too.
These results are in \autoref{tab:summintro}. For benchmarking, if both networks play all actions with equal probability in all games, the average MaxReg is 0.259 in \(2 \times 2\) games and 0.213 in \(3 \times 3\).

\begin{table}[h]
	\centering
	\footnotesize 
	
	\begin{minipage}{.46\linewidth}
		\centering
		\begin{tabular}{cccc}
			\multicolumn{4}{c}{\textbf{$\mathbf{2 \times 2}$ Games}} \\[2pt]
			\toprule
			& \multirow{2}{*}{\textbf{\begin{tabular}[c]{@{}c@{}}All\\[1pt] Games\end{tabular}}} & \multicolumn{2}{c}{\textbf{\# Pure Nash}} \\
			\cmidrule(lr){3-4}
			& & $\mathbf{0}$ & $\mathbf{\geq 1}$ \\
			\midrule
			\textbf{Relative Frequency} & \small 1.00  & \small 0.13 & \small 0.87 \\
			\midrule
			\textbf{Mean MaxReg} & \small 0.00 & \small 0.01 & \small 0.00\\[2pt]
			\textbf{} & \footnotesize (0.00) & \footnotesize (0.01) & \footnotesize (0.00) \\[4pt]
			\midrule
			\textbf{Benchmark} & \small 0.26 & \small 0.15 & \small 0.27\\[2pt]
			
			\bottomrule
		\end{tabular}
	\end{minipage}
	\hspace{0.04\linewidth}
	\begin{minipage}{.48\linewidth}
		\centering
		\begin{tabular}{cccc}
			\multicolumn{4}{c}{\textbf{ $\mathbf{3 \times 3}$ Games}} \\[2pt]
			\toprule
			& \multirow{2}{*}{\textbf{\begin{tabular}[c]{@{}c@{}}All\\[1pt] Games\end{tabular}}} & \multicolumn{2}{c}{\textbf{\# Pure Nash}} \\
			\cmidrule(lr){3-4}
			& & $\mathbf{0}$ & $\mathbf{\geq 1}$ \\
			\midrule
			\textbf{Relative Frequency}   & \small 1.00 & \small 0.21 & \small 0.79 \\
			\midrule
			\textbf{Mean MaxReg} & \small 0.01 & \small 0.03 & \small 0.01\\[2pt]
			\textbf{} & \footnotesize (0.03) & \footnotesize  (0.04) & \footnotesize (0.03)   \\[4pt]
			
			\midrule
			\textbf{Benchmark} & \small 0.21 & \small 0.19 & \small 0.22\\[2pt]
			
			\bottomrule
		\end{tabular}
	\end{minipage}

	\caption{Key results. Mean values over the test sets and by number of pure equilibria of the maximum normalised regret across players (MaxReg).  Standard deviation in parentheses. Benchmark is the mean maximum normalized regret arising from random play. Values are rounded.}
	\label{tab:summintro}
\end{table}

When restricting evaluation to games with a unique pure equilibrium, the networks play an \textit{exact} Nash in nearly all \(2\times2\) games and most \(3\times3\) ones. This ability is developed early in the training. In contrast, average MaxReg is larger for games with only one mixed equilibrium and remains positive after training. For instance, in the  Matching Pennies game, our trained network plays mixed strategy \((0.51,0.49)\). 

The difference in performance between the two scenarios is not surprising if one sees neural networks as programmable machines. Finding a pure equilibrium in a \( n\times n\) game requires making at most $O(n^2)$ payoff comparisons, whereas identifying a mixed equilibrium requires solving a system of equations, whose computational complexity grows non-polynomially with the number of actions (see \citet{Daskalakis2009}, \citet{Chen2007}).

The observation that the networks learn to play approximate Nash equilibria raises the question of which ones are being selected when multiple exist.  In almost all \(2 \times 2\) games and in 76\% of \(3 \times 3\) games with multiple equilibria, the networks select the risk-dominant equilibrium, as defined by (the linear tracing procedure of) \citet{Harsanyi1988}. Consistent with this selection criterion, mixed equilibria are only played in games with a single mixed equilibrium and in (nearly) symmetric games with multiple equilibria.
These findings reinforce arguments from evolutionary game theory (e.g., \citet{Kandori1993} and \citet{Young1993}) in favour of selecting risk-dominant equilibria in \(2 \times 2\) games and suggest some of these results may extend to populations of players jointly evolving a solution concept. On the other hand, they motivate caution toward viewing the refinement of \citet{Harsanyi1988} as the result of evolutionary forces in  \(3 \times 3\) games or larger.

To enhance the characterization, we tested the networks' behaviour against various axioms for single-valued solution concepts. We find that the same equilibrium is selected in a game when the role of players is swapped, as one expects if learning is independent of game-sampling. Additionally, the played equilibrium remains the same when payoffs are transformed in ways that do not alter best reply correspondences, or when payoffs are raised at the equilibrium point. The same invariance applies when the order of actions is permuted, indicating the networks do not exploit this opportunity to correlate their behaviour.  Lastly, we show that the same equilibrium is selected in \(2\times2\) games and in \(3 \times 3\) games built from those \(2\times2\) ones by adding a strictly dominated action for each player. Consistency between models of different sizes further demonstrates robustness of the learned behaviour.

A possible, unpleasant outcome of our approach is that it results in a trivial model, unable to generalise beyond the training set. Remarkably, this is not the case. The weights are used to build a function that approximates an algorithm for finding the relevant Nash equilibrium independently of payoff magnitudes. We confirm this claim by running two experiments. First, after training on games from a normalised space of utilities, we test on games drawn from a different space obtained by applying random positive affine transformations.  Second, we train the networks on selected \textit{quarters} of the set of all games, but evaluate them on the \textit{complement} set. In doing so, we also eliminate games that do not possess strategically equivalent alternatives in the training set. In both cases, the networks learn to behave close to the baseline models, as long as the space of included games is sufficiently rich.

Let us conclude this section with a discussion of two key informational assumptions: observability of the opponent's mixed strategies and payoffs. First, assuming our learners observe the mixed strategy of the opponents when computing their counterfactual loss implies we are treating those mixed strategies as pure strategies of the mixed-extension version of a standard \(n\times n\) game. In contrast to ordinal games with a finite number of actions, such mixed-extensions always possess a Nash equilibrium. We show in \autoref{app:mixed} that a substantial amount of learning takes place also when we let the counterfactual loss be evaluated on a realisation from the mixed strategy of the opponent.                   

Second, we assume complete information of the game being played at each iteration. This assumption is restrictive. However, the best we could hope for without information about the opponent's payoff would be for play to converge to the Bayes Nash equilibrium of the corresponding static Bayesian game. Indeed, if we do not provide the opponent's payoffs as input, because payoffs are independently drawn across players, the networks converge to playing the action that returns the highest expected payoff against a random action of the opponent.
Additionally, complete information about the game is common in laboratory experiments and can be appropriate in contexts where the adversary’s goal is sufficiently transparent, e.g., navigating traffic, where each driver’s objective of accident‐avoidance and efficient movement is relatively well understood by others. In such simple environments, it is plausible that across repeated interactions people build intuitive heuristics on how to play generic strategic situations, mirroring the across‐game machine learning that we study. 

The rest of the paper is organised as follows. In  \autoref{sec:neural_networks} we provide a formal description of our model. In \autoref{sec:baseline} we present our baseline specification and in \autoref{sec:results} our main results. In  \autoref{sec:dynamics}, we explore learning dynamics, highlighting the speed of convergence to Nash. In \autoref{sec:literature} we discuss the relevant literature and in \autoref{sec:conclusion} we conclude with additional thoughts and directions for future research. The robustness analysis and other technical details are in Appendices A-I. In particular, we show that results are robust to uniform sampling and remain largely unaffected by training models for larger games (\autoref{app:larger}) and when employing alternative model specifications (\autoref{app:architecture} and \autoref{app:sampling}).


\section{Training game playing neural networks}\label{sec:neural_networks}

We begin with some basic game theoretic definitions. Then, we define the neural network architecture we employ and explain how learning takes place.
\vspace{-0.15cm}


\subsection{Basic Definitions.}\label{sec:definitions}

An \(n \times n\) two-player game, \(U\), is a pair of real-valued \(n \times n\) matrices \((U^1,U^2)\). Element \((j,k)\) of \(U^i\) indicates the payoff of player \(i=1,2\) (row and column) when \(i\) plays action \(j\in \{1,\dots,n\}\) and the opponent chooses action \(k \in \{1,\dots,n\}\). We restrict attention to games where the payoff matrix of each player, denoted by \( \mathcal{S}_n \), is a point in \(\mathbb{R}^{n \times n}\) with mean zero and Frobenius norm \(n\). We denote by \(\mathcal{U}_{n \times n}=\mathcal{S}_n \times \mathcal{S}_n\) the set of all such games. The two normalisations imply that no two games in \(\mathcal{U}_{n \times n}\) have payoffs that are positive affine transformations of each other for any of the players. This choice reflects the view that preferences are the primitive of a game. More details are given in \autoref{app:space_of_games}.

Let \(\sigma^i \in \Delta^{n-1}\) denote a mixed strategy for player \(i=1,2\), represented as an \(n\times 1\) vector. Let  \(\sigma^{-i} \in \Delta^{n-1}\) denotes a mixed strategy for the opponent of \(i\). Given a strategy profile \(\sigma=(\sigma^1,\sigma^2)\), the expected payoff of \(i\) is given by \((\sigma^i)^\top  \hspace{0.1em} U^i \hspace{0.15em} \sigma^{-i}\). The \textit{regret} of \(i\) in game \(U\) for \(\sigma\), denoted \(R^i(U,\sigma)\), is the difference between the highest expected payoff delivered in game \(U\) by any pure strategy and that obtained by \(\sigma^i\) given \(\sigma^{-i}\). In matrix notation
\begin{equation} \label{regret} 
	R^i(U,\sigma)=\max_{j\in \{1,\dots,n\}}[U^i \sigma^{-i}]_j - (\sigma^i)^\top  U^i \sigma^{-i}\geq 0.
\end{equation}

A profile \(\sigma\) is a \textit{Nash equilibrium} of \(U\) if and only if  \(R^i(U,\sigma)=0\) for \(i=1,2\). A profile \(\sigma\) is an \(\epsilon\)-\textit{Nash equilibrium} (or simply an \(\epsilon\)-equilibrium) of  \(U\) if and only if  
\[
	\text{MaxReg}(U,\sigma)=\max_i \frac{R^i(U,\sigma)}{\max U^i-\min U^i}\leq \epsilon,
\] 
where \(\max U^i\) and \(\min U^i\) are the maximum and minimum payoffs of player \(i\) in game \(U\).


\subsection{Neural Networks} 

A \(n\)-action game-playing neural network for player \(i\) is a continuous and (almost everywhere) differentiable function \[f_n^i(\cdot \, ; w): \mathcal{U}_{n \times n} \rightarrow \Delta^{n-1},\] where \(w\) is a vector of trainable parameters (or weights). For each possible \(n\times n\) game, the neural network \(f_n^i\) outputs a mixed strategy for player \(i\), given the state of its parameters. As discussed, the network conditions its play on the entire game.

\begin{figure}[h]
	\centering
	\includegraphics{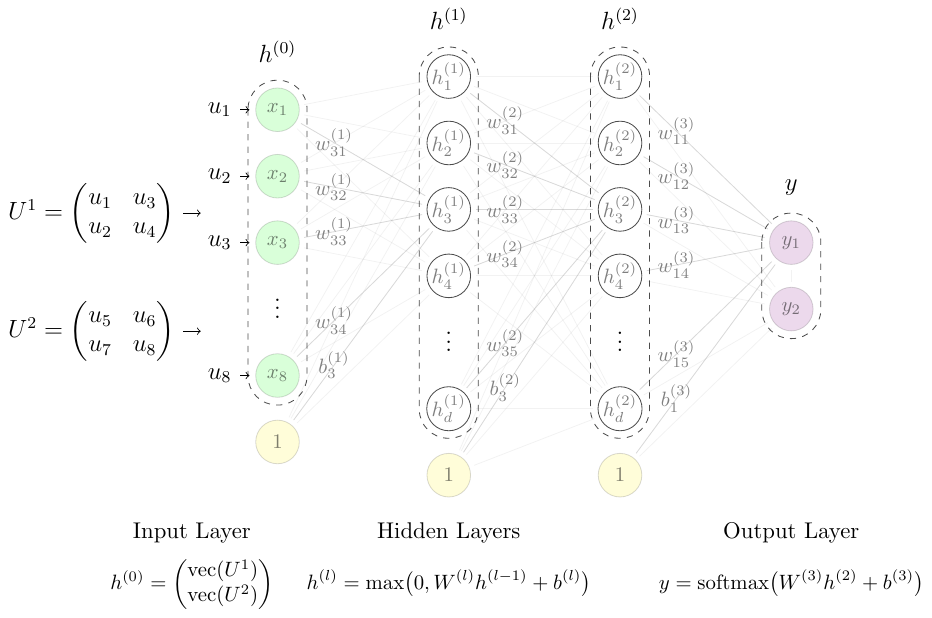}
	\caption{Game playing neural network with two hidden layers. Input nodes in \(h^{(0)}\) receive the eight payoffs of game \((U^1,U^2)\); nodes in the hidden layers, \(h^{(1)}, h^{(2)}\), and nodes in the output layer, \(y\) perform computations on the values stored in the previous layer. Softmax transforms the output into a probability distribution.}
		\label{fig:network}
\end{figure}

The functional form \(f^i_n\) is determined by the network architecture. We use a standard multi-layer feed-forward network composed of an input layer, \(L\ (\geq 1)\) hidden layers of dimension \(d\ (> 2n^2)\), and an output layer, all fully connected. The total number of parameters is \((L-1)d^2 + (2n^2+n+L)d + n\). Beyond being flexible enough to serve as a universal approximator, the advantage of the neural network's functional form is that it simplifies optimization by making differentiation over parameters parallelizable and computationally cheaper.  \autoref{app:network} contains additional details. An illustration is provided in \autoref{fig:network}.

\subsection{Training}

Training starts by initialising two independent networks with parameter vectors \(w^1_0\) and \(w^2_0\). Then, these are dynamically adjusted across a sequence of \(T\) periods. In each period \(t\), a batch \(B_t\) of \(B\) games is independently sampled from \(\mathcal{U}_{n \times n}\) according to a stationary distribution. Given parameters at time \(t\), denoted \(w^1_t\) and \(w^2_t\), the two networks generate play recommendations for each game \(U \in B_t\), for both the row player, \(f_n^1(U; w^1_t)\), and the column one, \(f_n^2(U; w^2_t)\). Finally, updating of the parameters for each network \(i\) takes place via stochastic gradient descent on a loss function defined by the square of the regret that players experience in the game.  

Formally, for \(t=0,\dots, T\) and \(i=1,2\), the parameters evolve according to
\[
	w^i_{t+1} = w^i_t - \eta_t \ \frac{1}{B}\sum_{U \in B_t} \nabla_{\!w}  R^i\left( U^i, f_n^i(U; w), \sigma_t^{-i}\right)^2  \! \big|_{w=w^i_t}
\]
where \(\sigma_t^{-i} = f_n^{-i}(U; w_t^{-i})\) is the strategy chosen by the opponent network in period \(t\) and \(\eta_t= \eta_0 \alpha^t\) is a learning rate decaying geometrically at rate \(\alpha \in (0,1]\).\footnote{Identical results would be achieved using a single neural network trained in self-play.} 

Training a neural network requires defining a differentiable loss function assigning a value to all available choices. In a game, this implies making a conjecture about the opponent's behaviour. As standard, we ground such a conjecture on the strategy chosen by the opponent at each iteration. In addition, letting the learner observe the opponent's mixed strategy, rather than an action realisation, removes noise and simplifies learning. As we acknowledged, this is not the standard choice in models of learning mixed strategy equilibria and one may want to view our mixed strategies as pure strategies of a larger mixed-extension game. Relaxation of this assumption is discussed in \autoref{app:mixed}.

A natural choice of the loss function would be to use the negative of the payoff: 
\[
	 - \ \pi^i(U^i,\sigma^{-i}; w^i)= - \ f^i_n(U; w^i)^\top  U^i \sigma^{-i}.
\]
However, this target is not strictly concave in $i$'s own action and attempting to maximise payoff directly results in non-convergent dynamics.\footnote{This is a well-known issue and it is common practice to regularise the loss function by introducing a strictly concave penalization \citep[e.g. see][Ch. 4]{FudenbergLevine98}. We adopt an analogous approach.}
To address this problem while keeping the same target, we employ a loss function based on squared regret: 
\[
	R^i(U,f^i_n(U; w^i),\sigma_{-i})^2.
\]
Squared regret has the same gradient that would govern the dynamics in the payoff maximisation problem, but scaled by twice the (positive) realised regret in the game:   
\[
	\nabla_w R^i(U,f^i_n(U;w),\sigma_{-i})^2 = - 2R^i (U,f^i_n(U; w^i),\sigma_{-i})\nabla_w \pi^i(U^i,\sigma^{-i}; w^i).
\]


\section{The baseline model}\label{sec:baseline}   

For \(2 \times 2\) (\(3 \times 3\)) games and for both players, we use a fully connected feedforward neural network with 8 (8) hidden layers of 256 (512) neurons, randomly initialised. The total number of parameters in the networks is 463,362 (1,849,859). For all models, the initial learning rate \(\eta_0\) is set to \(0.5\) and decays exponentially at rate \(\alpha=0.999999\). Hyperparameters are determined through a process of trial and error to balance computational cost, complexity and performance. \autoref{app:architecture} shows that findings are robust to alternative choices.   

We train our network players for \(2\times 2\) and \(3 \times 3\) games, sampling uniformly at random \(2^{30}\) (\(\approx\) 1 billion) games from \(\mathcal{U}_{2\times 2}\) and  \(\mathcal{U}_{3 \times 3}\), respectively.  The mean payoff in both classes of games is zero and its variance is one. Payoffs are in the interval \(\left[-\sqrt{n^2-1}, \sqrt{n^2-1} \, \right]\) for \(n\times n\) games. In \(2\times 2\) (\(3 \times 3\)) games, the average maximum individual payoff is \(1.292\) \((1.625)\) and the average minimum is \(-1.292\) \((-1.625)\). Batching is in groups of 128 (128) games for \(2\times 2\) (\(3\times 3\)) games. Uniform sampling of games is chosen for simplicity and implies that payoffs matrices are independent across players and degenerate games are sampled with probability zero. The sampling method is important because it determines the sets of games that networks are exposed to. However, the assumption of uniformity is not essential. As long as the set of games is rich enough and independence across periods is preserved, the trained behaviour of the network remains the same. Robustness to sampling and batching is explored in \autoref{app:sampling}.

The details of the neural network architecture, data generation, and the training process are summarised in the \autoref{tab:network} below for both the \(2\times 2\) and \(3\times 3\) models.    

\begin{table}[tbh]
	\footnotesize
	\centering
	\begin{tabular}{@{}l ll@{}}
 & \textbf{$\mathbf{2 \times 2}$ Games} & \textbf{$\mathbf{3 \times 3}$ Games} \\[2pt]
\toprule
\textbf{Network Sizes ($L \times d$)}     & $8 \times 256$           & $8 \times 512$ \\[2pt]
\textbf{Number of Weights}               & $463{,}362$              & $1{,}849{,}859$ \\[2pt]
\textbf{Sampling}                & $ \text{Uniform}(\mathcal{U}_{2 \times 2})$ & $\text{Uniform}(\mathcal{U}_{3 \times 3})$ \\[2pt]
\textbf{Number of Games ($K$)}           & $1{,}073{,}741{,}824$        & $1{,}073{,}741{,}824$ \\[2pt]
\textbf{Batch Size ($B$)}                & $128$                      & $128$ \\[2pt]
\textbf{Optimization Steps ($T$)}        & $8{,}388{,}608$         & $8{,}388{,}608$ \\[2pt]
\textbf{Learning rate ($\eta_0, \alpha$)}                    & $0.5$, $0.999999$ & $0.5$, $0.999999$ \\[2pt]
\bottomrule
\end{tabular}

	\caption{Summary of baseline models.}
	\label{tab:network}
\end{table}


\section{Main results}\label{sec:results}

In this section we report results from testing the trained baseline models on randomly generated (out-of-sample) test sets of \(2\times2\) and \(3\times3\) games, each containing \(2^{17}\) (131,072) games. We first provide evidence that the networks learn to avoid dominated and, more generally, non rationalizable strategies. Then, we show that they learn to play (approximate) Nash equilibria. After, we look at equilibrium selection. Our main observation is that in \(2\times 2\)  games with multiple equilibria, the networks learn to play the risk-dominant equilibrium. Finally, we consider out-of-distribution performance of the networks.

Throughout, we report the results here for a single pair of jointly trained models. We verify by running several simulations that there is little to no variance in behaviour arising from the initial seed, which regulates randomness in parameter initialisation and the realisation of the training set.  Results of these simulations are omitted and available upon request.


\subsection{Rationality} 

The networks learn to avoid strictly dominated actions. Looking at dominance solvable \(2\times2\) (\(3\times3\)) games, where at least one strictly dominated strategy exists for a player, the average probability mass placed on strictly dominated actions is \(0.002\)  \((0.007)\). Remarkably, the average mass on strictly dominated strategies goes down to nearly zero if we restrict attention to games where the maximal payoff loss from playing the dominated action is greater than \(0.1\). We interpret this finding as the network trading off its limited computation capacity between predicting the opponent's action and selecting an optimal response. 
The networks also learn that their opponents are rational and adjust their behaviour accordingly. In \(2\times2\) (\(3\times 3\)) dominance solvable games the mean mass placed on strategies that are not rationalizable is equal to \(0.003\) \((0.024)\).


\subsection{Nash play} 

\begin{figure}[tbh]
	\centering
	\includegraphics[width=0.495\textwidth]{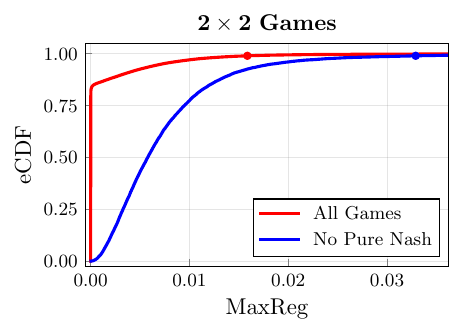}
	\includegraphics[width=0.495\textwidth]{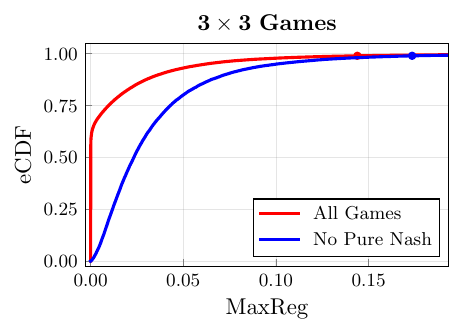} 
	\caption{Empirical cumulative density of MaxReg. Dots are placed in correspondence of the 99th percentile.}
	\label{fig:regret_cdf}
\end{figure}

\autoref{tab:summary} and \autoref{fig:regret_cdf} show that the networks learn to play an approximate Nash equilibrium in almost all games. Our key statistic is the maximum normalized regret across players experienced on average in the test set. In any given game \(U\), the maximum normalized regret across players (MaxReg) is the minimal \(\epsilon\) such that the networks are playing an \(\epsilon\)-equilibrium in \(U\).   In \autoref{tab:summary} we show the mean and standard deviation of the average MaxReg over the test set by number of Nash equilibria, for \(2 \times 2\) and \(3 \times 3\) games. \autoref{fig:regret_cdf} displays the empirical cumulative distribution of MaxReg achieved over the test set for all games and for games with only mixed Nash equilibria.

\begin{table}[tbh]
	\centering
	\footnotesize 
	\begin{tabular}{ccccc}
	\multicolumn{5}{c}{\textbf{Baseline -- $\mathbf{2 \times 2}$ Games}} \\[2pt]
	\toprule
	& \multirow{2}{*}{\textbf{\begin{tabular}[c]{@{}c@{}}All\\[1pt] Games\end{tabular}}}  & \multicolumn{3}{c}{\textbf{\# Pure Nash}} \\
	\cmidrule(lr){3-5} 
	& & $\mathbf{0}$ & $\mathbf{1}$ & $\mathbf{> 1}$ \\
	\midrule 
	
	\textbf{Relative Frequency} & \small 1.000 & \small 0.126  & \small 0.748 & \small 0.126 \\[2pt]
	\midrule
	\textbf{Mean MaxReg}  & \small 0.001 & \small 0.007 & \small 0.000 & \small 0.000 \\[2pt]
				 & \footnotesize (0.004) & \footnotesize (0.006) & \footnotesize (0.000) & \footnotesize (0.006) \\[4pt]
		\midrule
		\textbf{Benchmark} & \small 0.259 & \small 0.154 & \small 0.295 & \small 0.154\\[2pt]
		
	\bottomrule\\[5pt]
	
	\multicolumn{5}{c}{\textbf{Baseline -- $\mathbf{3 \times 3}$  Games}} \\[2pt]
	\toprule
	& \multirow{2}{*}{\textbf{\begin{tabular}[c]{@{}c@{}}All\\[1pt] Games\end{tabular}}}  & \multicolumn{3}{c}{\textbf{\# Pure Nash}} \\
	\cmidrule(lr){3-5} 
	& & $\mathbf{0}$ & $\mathbf{1}$ & $\mathbf{ > 1}$ \\
	\midrule 
	
	\textbf{Relative Frequency} & \small 1.000 & \small 0.213  & \small 0.580 & \small 0.207 \\[2pt]
	\midrule
	\textbf{Mean MaxReg}  & \small 0.012 & \small 0.034 & \small 0.006 & \small 0.006 \\[2pt]
 & \footnotesize (0.030) & \footnotesize (0.036) & \footnotesize (0.023) & \footnotesize (0.031) \\[4pt]
 \midrule
 \textbf{Benchmark} & \small 0.213 & \small 0.188 & \small 0.230 & \small 0.189\\[2pt]
 
	\bottomrule
\end{tabular}

	\caption{ Average MaxReg. Mean values of MaxReg over test sets of \(2^{17}\) randomly generated games and across subgroups defined by the number of pure Nash equilibria. Standard deviation in parentheses.  Benchmark is the mean maximum normalized regret arising from random play. Values are rounded.}
	\label{tab:summary}
\end{table}

In \(2\times 2\) games, the networks obtain an average MaxReg of \(0.001\), compared to an average MaxReg from fully random play of \(0.259\). By looking at the distribution of the MaxReg obtained across the two players, we see the networks are playing an \(\epsilon\)-Nash equilibrium with \(\epsilon\leq 0.016\) in \(99\%\) of games. A better performance is achieved in games with only one pure Nash equilibrium where the average MaxReg is below \(0.001\) and an \textit{exact} (up to machine precision) Pure Nash is played around \(85\%\) of times. On the other hand, in games with a unique equilibrium in mixed strategies the average MaxReg is \(0.007\). 

In \(3 \times 3\) games the performance of the network degrades but learning still takes place. Players achieve an average MaxReg of  \(0.012\), compared to a \(0.213\) benchmark from random play. They are playing an \(\epsilon\)-Nash equilibrium with \(\epsilon<0.144\) \((0.062)\) in \(99\%\) \((95\%)\) of games.  Average MaxReg goes down to \(0.006\) in games with a unique pure equilibrium. As was the case for \(2\times 2\) games, but to a greater extent, the networks have difficulties accurately learning to play mixed strategy equilibria. In the case of games with only one mixed equilibrium, the average MaxReg reaches \(0.034\).


\subsection{Behaviour in a representative subspace of 2x2 games.} 

\begin{figure}[tbh]
	\centering
	\includegraphics[width=1\textwidth]{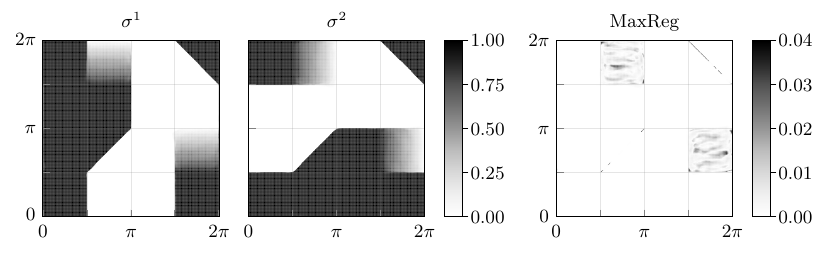}
	\caption{Heatmaps over the strategic subspace of \(\mathcal{U}_{2\times 2}\), parameterized by angular coordinates \citep{jessie2015}.  The first two plots show the probabilities assigned to the first action by each network, \((\sigma^1)_1\), \((\sigma^2)_1\). The third plot shows the corresponding maximum normalized regret across agents, \(\mathrm{MaxReg}\).}
	\label{fig:strategic_subspace}
\end{figure}

 In \autoref{fig:strategic_subspace}, we show the networks’ strategies and \(\mathrm{MaxReg}\) on a measure‐zero subspace of \(\mathcal{U}_{2\times 2}\), denoted by \(\mathcal{U}^{\mathrm{N}}_{2\times2}\), that captures all possible strategic interactions \citep{jessie2015}. Games in this subspace are unique up to best-reply structure equivalence and can be represented as points in a \([0,2\pi)\times[0,2\pi)\) square. The first coordinate of each plot in \autoref{fig:strategic_subspace} determines Player 1’s payoff matrix and the second determines that of Player 2. The space is partitioned into a \(4\times4\) grid with cells numbered top-to-bottom and left-to-right. Cells (1,2) and (3,4) consist of games with a unique mixed-strategy equilibrium, while cells (3,2) and (1,4) consist of games with three equilibria; all remaining cells correspond to games with a unique pure-strategy equilibrium. Moreover, Player 1 has a dominant action in games located along columns 1 and 3, while Player 2 has a dominant action in games located along rows 1 and 3. Finally, symmetric games are located along the \(45^\circ\) diagonal. Details are provided in \autoref{app:space_of_games}.

Beyond confirming that pure equilibria are approximated accurately, this visualization provides two additional insights. First, regret is higher in regions where the selected equilibrium is changing. Second, examining mixed equilibria further reveals that, as one would expect, changes in a player’s own payoffs do not affect their equilibrium mixing when mixing remains the only equilibrium, but they alter the equilibrium mixing of the opponent.


\subsection{Equilibrium selection} 

To gain insight on the equilibrium selection, we look further at games in our test set with more than one Nash equilibrium. Games with multiple equilibria represent \(12.6\%\) of all \(2 \times 2\) games and  \(20.7\%\) of all \(3\times 3\) games in the test sets.

For each of the games with multiple equilibria in our test sets, we compare the strategy profile produced by the networks to each Nash equilibrium of the game and we select the equilibrium with the smallest total variation distance.\footnote{Two strategies \(\sigma^i\) and \(\tilde {\sigma}^{i}\) have total variation distance equal to \(TV(\sigma^i,\tilde {\sigma}^{i})=\frac{1}{2} \sum_j | \sigma^i_j -\tilde {\sigma}^{i}_j |\). The total variation distance of profile \((\sigma^1,\sigma^2)\) from another profile \((\tilde {\sigma}^{1} ,\tilde {\sigma}^{2} )\) is \(\max \{TV(\sigma^1,\tilde {\sigma}^{1}), TV(\sigma^2,\tilde {\sigma}^{2})\}\)} We then classify this closest equilibrium according to two binary criteria: whether it is risk dominant and whether it is utilitarian (i.e., maximises the sum of payoffs). Finally, for all games with multiple equilibria and for all such games with a payoff dominant equilibrium (i.e.,  unique Pareto optimal equilibrium), we compute the distribution over the four possible categories for our closest equilibrium.\footnote{When it exists, the unique Pareto optimal equilibrium is the unique utilitarian equilibrium.} The results are in \autoref{tab:selection}.

\begin{table}[h]
	\centering
	\footnotesize
\begin{minipage}{0.45\linewidth}
\centering
\begin{tabular}{ccc|c}
	\multicolumn{4}{c}{\textbf{$\mathbf{2 \times 2}$ Games -- All Games}} \\[2pt]
	\toprule
	& \multirow{2}{*}{\textbf{Utilitarian}} & \textbf{Not} & \textbf{} \\
	& & \textbf{Utilitarian} & \textbf{} \\
	\midrule
	\textbf{Risk} & \small 0.662 & \small 0.328 & \small 0.990 \\
	\textbf{Dominant} & \tiny \{0.993\} & \tiny [0.984] & \small  \\[4pt]
	\textbf{Not Risk} & \small 0.003 & \small 0.006 & \small 0.010 \\
	\textbf{Dominant} & \tiny [0.010] & \tiny \{0.007\} & \small  \\[4pt]
	\midrule
	\textbf{} & \small 0.665 & \small 0.335 & \\
	\bottomrule
\end{tabular}
\end{minipage}
\hspace{0.05\linewidth}
\begin{minipage}{0.45\linewidth}
\centering
\begin{tabular}{ccc|c}
	\multicolumn{4}{c}{\textbf{$\mathbf{2 \times 2}$ Games -- Payoff Dominant Exists}} \\[2pt]
	\toprule
	& \textbf{Payoff} & \textbf{Not Payoff} & \textbf{} \\
	& \textbf{Dominant} & \textbf{Dominant} & \\
	\midrule
	\textbf{Risk} & \small 0.710 & \small 0.280 & \small 0.990 \\	
	\textbf{Dominant} & \tiny \{0.996\} & \tiny[0.976] & \small  \\[4pt]
	\textbf{Not Risk} & \small 0.004 & \small 0.005 & \small 0.010 \\
	\textbf{Dominant} & \tiny [0.015] & \tiny\{0.004\} & \small  \\[4pt]
	\midrule
	\textbf{} & \small 0.715 & \small 0.285 & \\
	\bottomrule
\end{tabular}
\end{minipage}

\vspace{15pt}

\begin{minipage}{0.45\linewidth}
\centering
\begin{tabular}{ccc|c}
	\multicolumn{4}{c}{\textbf{$\mathbf{3 \times 3}$ Games -- All Games}} \\[2pt]
	\toprule
	&  \multirow{2}{*}{\textbf{Utilitarian}} & \textbf{Not} & \textbf{} \\
	& & \textbf{Utilitarian} & \textbf{} \\
	\midrule
	\textbf{Risk} & \small  0.479 & \small 0.276 & \small 0.755 \\
	\textbf{Dominant} & \tiny \{0.807\} & \tiny[0.678] & \small  \\[4pt]
	\textbf{Not Risk} & \small 0.105 & \small 0.140 & \small 0.245 \\
	\textbf{Dominant} & \tiny [0.259] & \tiny\{0.193\} & \small  \\[4pt]
	\midrule
	\textbf{} & \small 0.584 & \small 0.416 & \\
	\bottomrule
\end{tabular}
\end{minipage}
\hspace{0.05\linewidth}
\begin{minipage}{0.45\linewidth}
\centering
\begin{tabular}{ccc|c}
	\multicolumn{4}{c}{\textbf{$\mathbf{3 \times 3}$ Games -- Payoff Dominant Exists}} \\[2pt]
	\toprule
	& \textbf{Payoff} & \textbf{Not Payoff} & \textbf{} \\
	& \textbf{Dominant} & \textbf{Dominant} & \\
	\midrule
	\textbf{Risk} & \small 0.524 & \small 0.236 & \small 0.761 \\
	\textbf{Dominant} & \tiny\{0.822\} & \tiny [0.653] & \small  \\[4pt]
	\textbf{Not Risk} & \small 0.104 & \small 0.135 & \small 0.239 \\
	\textbf{Dominant} & \tiny[0.287] & \tiny \{0.178\} & \small  \\[4pt]
	\midrule
	\textbf{} & \small 0.628 & \small 0.372 & \\
	\bottomrule
\end{tabular}
\end{minipage}
	\caption{Equilibrium selection in games with multiple equilibria. Tables on the left consider all such games (16564 for \(2\times 2\) and 38591 for \(3\times 3\) games, respectively),  while tables on the right are restricted to games where a payoff-dominant equilibrium exists (8358 and 20206 games, respectively). Tables on the left (right) contain the sample frequencies of the network play for the four possible outcomes: utilitarian (payoff dominant) and risk dominant, risk dominant but not utilitarian (not payoff dominant), not risk-dominant but utilitarian (payoff dominant), not risk dominant and not utilitarian (not payoff dominant). Numbers in curly (square) brackets report frequencies conditional on the risk dominant equilibrium being (not being) utilitarian or payoff dominant.}
	\label{tab:selection}
\end{table}    

Remarkably, in 99\% of \(2 \times 2\) games with multiple equilibria the networks select the  \citet{Harsanyi1988} risk-dominant equilibrium. Non degenerate \(2 \times 2\) games have either a unique equilibrium or two pure equilibria and a mixed one. In such games, the risk dominant equilibrium is the pure equilibrium that minimises the product of the losses from individual deviations from it (or the mixed one if products are identical).\footnote{In  \(
	\left(\begin{smallmatrix}
	A,a \ &  B,b \\
	C,c \ & D,d
	\end{smallmatrix}\right)
	\) with equilibria at  \((A,a)\) and \((D,d)\) the  products of losses are \((A-C)(a-b)\) and \((D-B)(d-c)\).}  Since the space where the products of deviations are equal has measure zero in the space of all games, mixed strategy are almost never played by the networks in games with multiple equilibria.

Behaviour is more nuanced in  \(3\times 3\) games. The risk-dominant equilibrium computed according to the \citet{Harsanyi1988} linear tracing procedure (generalising the definition of risk-dominant equilibria given for \(2\times 2\) games) is selected only 76\% of the time. This figure goes down to 68-65\% (raises to 80-82\%) when the risk dominant equilibrium differs (coincides) with the utilitarian or payoff dominant. Even when the risk-dominant is not selected, the utilitarian or payoff-dominant equilibria are only selected about 40\% of times. Hence, the networks appear to often play Pareto dominated equilibria. The conclusion that mixed strategies are rarely played with multiple equilibria carries on to \(3\times 3\) games.


\subsection{Behavioural Axioms}

To enhance the characterization and provide data for an equilibrium selection theory in \(3\times 3\) games, we test adherence to a number of behavioural axioms. We find that the selection operated by the networks satisfies: symmetry  (i.e., the two networks select the same profile of strategies whenever the role of players is swapped), independence from strategically irrelevant actions (i.e.,  the networks play the same strategy in any two games where one is obtained by adding strictly dominated actions to the other), equivariance (i.e., the two networks select the same profile of strategies whenever two games differ only because the order of actions has been reshuffled), monotonicity (i.e., for all games where the network select a pure equilibrium, the pure equilibrium is still selected if we raise the payoff of players at the equilibrium point), and invariance to payoff transformations that preserve the best reply structure. The axioms, the tests performed and the numerical results are described in \autoref{app:axioms}. 

Symmetry implies the networks play symmetric equilibria in symmetric games.  Independence from strategically irrelevant actions implies coherence between the output of models of different sizes. Equivariance suggests the networks do not learn to correlate their strategy to the strategically irrelevant order of actions, but treat two equivariant games as the same. Monotonicity and invariance to payoff transformations that preserve the best reply structure characterise, together with equivariance, risk-dominant equilibria in \(2\times 2\) games \citep[see][Ch. 3]{Harsanyi1988}. Our results indicate these carry over to \(3\times 3\) games, albeit the evidence is less compelling.


\subsection{Out-of-distribution Performance} 

We now demonstrate, by means of two different exercises, that the trained neural networks perform well even for games outside of the support of the distribution of games on which they have been trained, as long as the training set is sufficiently rich.  Evaluating the models on data outside the support of the training distribution is a test of their generalization capabilities. If the networks were simply overfitting to the specific instances seen during training, say, by storing a finely partitioned grid of training examples, they would likely perform poorly when confronted with games whose payoff matrices do not fall within the training support. In contrast, if the networks have learned a robust, underlying heuristic (or the approximation of an algorithm) for computing Nash equilibria, then they should be able to deliver low-regret strategies even on games different from those encountered during training.   

First, we evaluate the networks on a set of \(2 \times 2\) games generated by applying random positive affine transformations to the payoff matrixes of the games in the test set for the baseline model.\footnote{Each game in the training set, \( (U^1, U^2) \), is transformed into \( (\alpha^1 U^1 + \beta^1,\, \alpha^2 U^2 + \beta^2) \), where \(\alpha^i\) is drawn uniformly from \([1,n]\) and \(\beta^i\) is drawn uniformly from \([-n,n]\). We restrict \(\alpha^i \geq 1\) to avoid generating games where a player is nearly indifferent between strategy profiles.} As discussed, the support of the training set contains no two games that can be generated from each other by means of affine transformations of individual payoffs. Hence, the performed transformations bring us outside the support of the training set. The results are in \autoref{tab:outofsample}. The average MaxReg is close to that obtained in the original test set and the total variation distance between the behaviour of the trained model and that of the baseline is small, albeit with significant variance.  

\begin{table}[tbh]
	\footnotesize
	\centering
	
\begin{tabular}{ccccc}
	\multicolumn{5}{c}{\textbf{Out-of-distribution -- $\mathbf{2 \times 2}$ Games}} \\[2pt]
	\toprule
	& \multirow{2}{*}{\textbf{\begin{tabular}[c]{@{}c@{}}All\\[1pt] Games\end{tabular}}}  & \multicolumn{3}{c}{\textbf{\# Pure Nash}} \\
	\cmidrule(lr){3-5} 
	& & $\mathbf{0}$ & $\mathbf{1}$ & $\mathbf{> 1}$ \\
	\midrule 
	
	\textbf{Relative Frequency} & \small 1.000 & \small 0.126  & \small 0.748 & \small 0.126 \\[2pt]
	\midrule
	\textbf{Mean MaxReg}  & \small 0.005 & \small 0.022 & \small 0.001 & \small 0.010 \\[2pt]
				 & \footnotesize (0.032) & \footnotesize (0.020) & \footnotesize (0.007) & \footnotesize (0.083) \\[4pt]
	\textbf{Mean DistBaseline}  & \small 0.027 & \small 0.072 & \small 0.019 & \small 0.030 \\[2pt]
				 & \footnotesize (0.122) & \footnotesize (0.114) & \footnotesize (0.114) & \footnotesize (0.162) \\[4pt]
	\bottomrule
\end{tabular}

	\vspace{-8pt}
	\caption{Evaluation on affinely transformed games. Mean maximum normalized regret across players (MaxReg) and mean total variation distance to the baseline model (DistBaseline) over \(2^{17}\) affinely transformed \(2\times 2\) games; standard deviation in parentheses. Values are rounded.}
	\label{tab:outofsample}
\end{table}

\begin{figure}[tbh]
	\centering
	\includegraphics{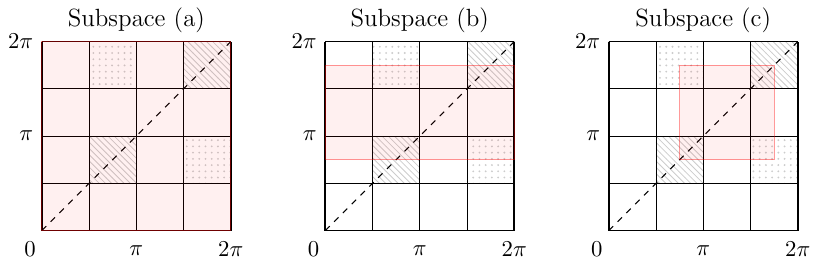}
	\caption{Strategic interactions included in each subspace as projected into \( \mathcal{U}^N_{2\times 2} \).}
	\label{fig:sss}
\end{figure}

Second, we train our baseline networks for \(2 \times 2\) games on three subspaces of \( \mathcal{U}_{2\times 2} \), each covering one quarter of the full space. These subspaces differ in the types of strategic interactions included in the training set. The first subspace contains a quarter of the whole space, but still includes all possible strategic interactions that arise in \(2 \times 2\) games.
The second subspace excludes all games where action 1 is dominant for the column player, and removes half of the games with three equilibria or a unique mixed-strategy equilibrium. The third subspace is more restrictive. It excludes all games where action 1 is dominant for either player as well as most games with a unique mixed or multiple equilibria. \autoref{fig:sss} shows the strategic interactions included in each subspace by projecting the games into \( \mathcal{U}_{2\times 2}^N \), the space of strategically distinct games. A formal description of the subspaces given in \autoref{app:subspaces}.

For each scenario, we evaluate the models on the complement of their training subspace. That is, on games never included and often distant (in the space of games) from any they had encountered during training. Results are in \autoref{tab:network_scenarios1}. 

\begin{table}[h]
	\footnotesize
	\centering
	\begin{tabular}{lccccc}
	&\multicolumn{5}{c}{\textbf{Training on subspaces -- $\mathbf{2 \times 2}$ Games}} \\[2pt]
   	 \cmidrule[1pt](lr){2-6}
	&& \multirow{2}{*}{\textbf{\begin{tabular}[c]{@{}c@{}}All\\[1pt] Games\end{tabular}}} & \multicolumn{3}{c}{\textbf{\# Pure Nash}} \\
	\cmidrule(lr){4-6}
	&& & $\mathbf{0}$ & $\mathbf{1}$ & $\mathbf{> 1}$ \\
	\cmidrule[0.6pt](lr){2-6}
	&\textbf{Relative Frequency} & \small 1.000  & \small 0.126 & \small 0.748 & \small 0.126 \\[4pt]
	\cmidrule[0.6pt](lr){2-6}
	 \multirow{4}{*}{\small \textbf{Subspace (a)}}&\textbf{Mean MaxReg} & \small 0.007 & \small 0.028 & \small 0.004 & \small 0.008 \\
							& & \footnotesize (0.025) & \footnotesize (0.030) & \footnotesize (0.018) & \footnotesize (0.043) \\[4pt]
		&\textbf{Mean DistBaseline} & \small 0.055 & \small 0.110 & \small 0.044 & \small 0.061 \\
							& & \footnotesize (0.177) & \footnotesize (0.153) & \footnotesize (0.172) & \footnotesize (0.210) \\[4pt]
	 \cmidrule[0.6pt](lr){2-6}
	 \multirow{4}{*}{\small \textbf{Subspace (b)}}&\textbf{Mean MaxReg} & \small 0.036 & \small 0.061 & \small 0.033 & \small 0.032 \\
							& & \footnotesize (0.099) & \footnotesize (0.077) & \footnotesize (0.102) & \footnotesize (0.096) \\[4pt]
		&\textbf{Mean DistBaseline} & \small 0.158 & \small 0.260 & \small 0.143 & \small 0.143 \\
							& & \footnotesize (0.335) & \footnotesize (0.306) & \footnotesize (0.336) & \footnotesize (0.338) \\[4pt]
	 \cmidrule[0.6pt](lr){2-6}
	 \multirow{4}{*}{\small \textbf{Subspace (c)}}&\textbf{Mean MaxReg} & \small 0.206 & \small 0.513 & \small 0.186 & \small 0.021 \\
							&& \footnotesize (0.327) & \footnotesize (0.336) & \footnotesize (0.316) & \footnotesize (0.120) \\[4pt]
		&\textbf{Mean DistBaseline} & \small 0.411 & \small 0.696 & \small 0.370 & \small 0.374 \\
							& & \footnotesize (0.466) & \footnotesize (0.176) & \footnotesize (0.480) & \footnotesize (0.482) \\[4pt]
	 \cmidrule[1pt](lr){2-6}
\end{tabular}
	\caption{Evaluation on games outside the training subspaces. Mean maximum normalized regret across players (MaxReg) and mean total variation distance to the baseline model (DistBaseline) over test sets of \(\approx3\times2^{15}\) games. Standard deviation in parentheses. Values are rounded.}
	\label{tab:network_scenarios1}
\end{table}

We find that in the first two scenarios the networks perform remarkably well in playing all games even when training on a subset of the space. Moreover, the behaviour learned on a quarter of all games is the same as that learned by the baseline models, as evidenced by the small total variation distance between the two set of models. Incidentally, this observation also demonstrates robustness of learning to the game sampling process.  

However, we observe lack of learning for models trained on the third scenario, in which we eliminate most of the games with a unique mixed equilibrium and most of those with multiple equilibria. This negative result reminds us that, while learning can take place even within a small subset of games, the choice of subset of games in which training takes place is crucial.


\section{Learning dynamics}\label{sec:dynamics}

\begin{figure}[h]
	\centering
	\includegraphics[width=0.495\textwidth]{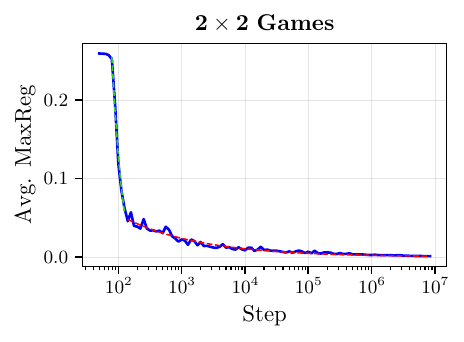}
	\includegraphics[width=0.495\textwidth]{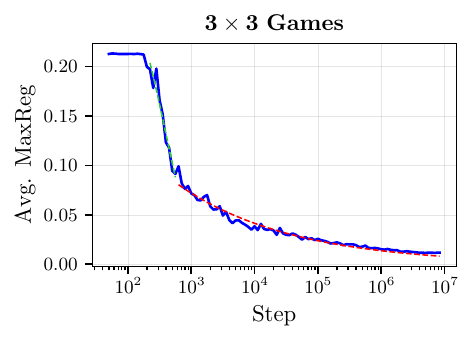}
	\caption{Learning curves for \(2 \times 2\) and \(3 \times 3\) models. The \(y\)-axis shows the maximum normalized regret across periods, averaged over the test set. The \(x\)-axis uses a base-10 logarithmic scale.}
	\label{fig:learning_curve}
\end{figure}

\begin{figure}[h]
	\centering
	\includegraphics[width=0.495\textwidth]{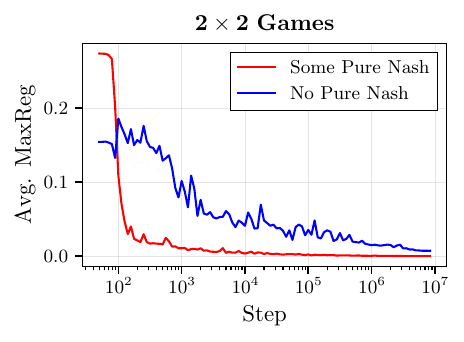}
	\includegraphics[width=0.495\textwidth]{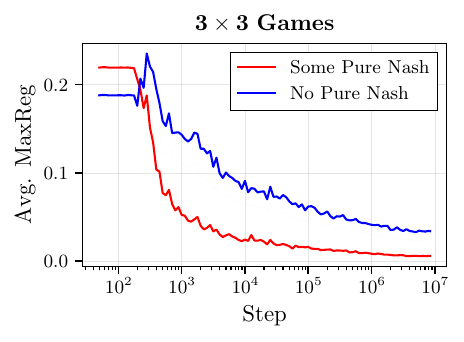}
 	\caption{Evaluation of average maximum normalized regret across players at logarithmically sampled periods during training, for games with only mixed equilibria and games with at least one pure equilibrium. The \(x\)-axis uses a base-10 logarithmic scale.}
	\label{fig:learning}
\end{figure}

In this section, we briefly discuss the learning dynamics for our baseline scenario. In \autoref{fig:learning_curve} we display the learning curves for \(2 \times 2\) and \(3 \times 3\) games. Specifically, we plot the maximum normalised regret achieved on average over the test set by the two networks on the \(y\)-axis, against the periods of training performed, from 1 to \(T\), on the \(x\)-axis in a base-10 logarithmic scale. In \autoref{fig:learning} we display the learning curves with evaluation of the maximum normalized regret performed separately for games that have some pure Nash equilibrium and games that have only mixed Nash equilibria.

Inspection of the learning curves suggests the following observations. First, there is an initial and very short period of little or no learning, lasting around 70 (100) optimization steps. Second, the networks experience a short phase of exponential learning, lasting around 70 (110) optimization steps. In this phase, most of the learning to play pure Nash equilibria takes place. The average regret goes down from the random-play benchmark to roughly \(0.025\) in less than 250,000 games of training. The best-fit exponential curve, depicted in dashed green in both graphs in \autoref{fig:learning_curve}, has an exponential decay rate of roughly 0.033 for the \(2\times 2\) models and of 0.002 for \(3\times 3\). Third, learning in the last and longest phase follows a power law. Playing of pure Nash equilibria is perfected while playing of mixed equilibria improves steadily. The best-fit power curve, depicted in red in the graphs, has a power-law exponent of roughly \(-0.360\) for \(2\times 2\) models and of \(-0.240\) for \(3\times 3\). Extrapolation from the power law indicates that \(55\) billion further periods of training may be required for the \(3\times 3\) models to bring down the average maximum normalized regret to the performance of \(0.001\) achieved in \(2\times 2\) games. 

Finally, while we observe there is a clear drift toward the equilibrium point of playing Nash in almost all games, there could be spiralling around such point with smaller and smaller amplitude. This suggests that time-averaging of the iterates is likely to deliver better performance in practice. Moreover, a formal convergence result for long-run time-average of the iterates might be easier to obtain than for last-time iterates.


\section{Literature Review}\label{sec:literature}

This paper has computational precursors.  To our knowledge, the idea of using a regret-based approach to train a neural network is first discussed in \citet{Marchiori2008} for a set of fixed games. But closest to ours is the work of \citet{Spiliopoulos2011,Spiliopoulos2012}. He used an adversarial approach to jointly train a population of neural networks to minimise the distance from best responding using randomly generated \(2 \times 2\) and \(3 \times 3\) games. While we share a similar setup, he focuses on pure equilibria and his results do not support the hypothesis of convergence to Nash. For \(3\times 3\) games with a unique pure equilibrium, he finds a 62\% frequency of (ex-post) Nash play, against \(96\%\) in our case using his benchmark.
Likely due to the limited availability of computational power at that time, the author concludes that neural networks are learning behavioural heuristics different from Nash.   

In their pioneering work, \citet{Sgroi2009} trained a single neural network to identify Nash equilibrium in \(3\times 3\) games. In contrast to us, the network was trained via supervised learning on random games with only one Nash equilibrium, with the aim to minimise the output's distance from the Nash equilibrium strategy.  \citet{Sgroi2009} are not preoccupied with the result of an adversarial learning process, but focus on the ability of the network to find the Nash equilibrium. While they concluded that neural networks would be unlikely to be able to learn Nash, recent engineering research on the learnability of equilibria by means a single neural network, e.g, \citet{Liu2024} and \cite{Zhijian2023}, reaches a conclusion in line with our finding.\footnote{A different machine learning angle is pursued in some recent work where neural networks, \citet{Hartford2016}, or related algorithms, \citet{Fudenberg2019}, are trained on existing experimental data with the aim of predicting human behaviour.} 

A handful of economics papers have explored theoretical models where learning takes place through a sequence of randomly generated games.\footnote{For a survey of the classic literature on learning in fixed games we refer to  \citet{FudenbergLevine98}.} In a seminal contribution, \citet{LiCalzi1995} studied fictitious-play dynamics. In his model, agents best respond upon observing which game they are playing, but beliefs over an opponent's behaviour are formed by pooling their actions in all past games. 
 \citet{Steiner2008} models the play of games never seen before by equipping the space from which games are drawn with a similarity measure. Players best respond to their learned belief about the behaviour of opponents, which is determined by a weighted average of behaviour on past play based on the measure of closeness between games. \citet{Mengel2012} studies players engaging in adversarial reinforcement learning over both how the best partition a finite space of games, subject to a convex cost of holding finer partitions, and which behaviour to adopt in each element of the partition. Her approach with its endogenous partitioning of games is close in spirit to ours.
However, her assumption that the set of possible games is finite allows learning to take place game-by-game when partitioning costs are small, which is in contrast to both \citet{Steiner2008} and us. 

There is some work on uncoupled learning in varying games. \citet{Cardoso2019} and \citet{Zhang2022} study repeated play in two-player zero-sum games where payoffs change over time, proposing online algorithms to minimise different regret notions. \citet{Harris2022} applied meta-learning across sequences of similar games, also using feedback within each game to adapt online learners, aiming to leverage cross-game similarity for faster convergence. These approaches contrast with our framework, where the dynamic is coupled and every new game is observed beforehand. Consequently, if games vary randomly and are played only once, as in our setting, the online adaptation core to the aforementioned approaches offers little benefit.     

The networks provide a unique recommendation for any possible game as a result of a competitive process. An analogous approach has been followed by others with different methodologies.  In \citet{Selten2003}, competing groups of students were asked to write programs to play any game. The programs were tested against each other with randomly generated games and feedback was provided. Programs were updated by the students and tested again, and the process was repeated for five iterations. The software produced in the course of a semester failed to compute equilibrium in games with only one mixed strategy but achieved 98\% success in choosing a pure Nash in dominant solvable games. With multiple pure equilibria, the utilitarian one was favoured. 

Recently, \citet{Lensberg2021} has pursued a related idea, but using genetic algorithms rather than students. A randomly mutating population of rules to play games compete at each iteration and more successful ones reproduce more. While \citet{Lensberg2021}  agree with us regarding the selection of the risk-dominant equilibrium in \(2\times 2\) games, both they and \citet{Selten2003} conclude,  in contrast to us, that the average heuristic at convergence is not, in general, a refinement of Nash.

Other authors have also studied the evolution of heuristics to play games. \citet{Samuelson2001} characterises the choice, subject to complexity costs, of a finite automaton for playing random games chosen from three classes.  \citet{Stahl1999} proposes a theory where a set of exogenous behaviour rules is reinforced based on their success. Relatedly,  \citet{Germano2007} considers evolutionary dynamics for a set of exogenous heuristics, with selection based on the success over a distribution of games. Among other features, our learning differentiates from \citet{Samuelson2001} as the set of games we deal with is large, and from \citet{Stahl1999} and \citet{Germano2007} because the set of possible rules of play is unrestricted.

Finally, our work complements (and is supported by) the sparse experimental evidence that empirically demonstrates the ability of human subjects to extrapolate across games. The experiments of \citet{Cooper2003}, \citet{Grimm2009}, \citet{Devetag2005}, \citet{Marchiori2008} and \citet{Marchiori2021} all show that strategic abilities can be learned and transferred to games that were not faced in the learning stage.


\section{Conclusions}\label{sec:conclusion}

In this paper, we show that two neural networks trained adversarially to maximise their expected payoff across sequences of fully observed randomly generated static games learn to select strategies corresponding to approximate Nash equilibria. This result reinforces the learning foundation of Nash equilibrium as a solution concept by showing that such strategies can be learned by analogy. 

Several promising directions remain for future research. From a theoretical perspective, our numerical results suggest it may be feasible to prove almost-sure convergence to Nash equilibria (or arbitrary $\epsilon$-Nash), either through the last iterate or through time-averaged strategies, for the gradient-based, regret-minimising discrete processes considered here. A natural approach would consist in analyzing parameter evolution for universal functional approximators through stochastic approximation techniques.     

Regarding computational extensions, we believe two directions are particularly compelling. First, verifying whether convergence to Nash equilibrium persists in settings involving more than two players. Second, making sure learning also occurs when opponent actions are explicitly sampled from their mixed strategies. The second task will require departing from a squared gradient loss, which biases the choice of optimal mixed strategy when the opponent's behaviour is uncertain. 

Further numerical work could also consider different classes of games. Studying scenarios involving incomplete information, such as Bayesian games, where payoffs are observed imprecisely, appears promising. Similarly, extending our approach to dynamic settings, where Nash equilibrium lacks strong learning foundations, represents an interesting research direction. Both extensions above face challenges due to increased dimensionality in game representations, necessitating substantial computational resources. Consequently, the use of abstraction techniques that enable agents to manage complexity will also be warranted.

Finally, while understanding strategic decision-making processes in machine intelligence has practical importance, we hope our results will stimulate experimental research into human strategic reasoning and learning. We have a few possible directions in mind. 

First, our findings raise experimental questions regarding parallels between neural network dynamics and human learning processes. Specifically, we observed that neural networks exhibit varying performance depending on the games they are trained on; future work might test whether similar comparative statics emerge in human learning scenarios. Investigating conditions under which neural-network-based models replicate human adaptive behavior would thus be valuable.

Second, our work could serve to gain insight on human strategic decision-making in environments requiring rapid responses.\footnote{\citet{Kahneman2011} distinction between fast and slow thinking extends to strategic decision-making contexts; see, for example, \citet{Rubinstein2016}'s analyses on response times.} It is conceivable that individuals learn to intuitively approximate Nash equilibrium behaviour even when encountering entirely novel interactions. Although testing this conjecture rigorously may be challenging within traditional laboratory experiments, practical evidence from rapid-decision domains, such as sports or competitive gaming, suggests that human strategic decision-making often relies heavily on intuition \citep[e.g.,][]{Walker2001,Chiappori2002, PalaciosHuerta2003}. 

Finally, we would find it valuable to benchmark our models against game playing humans or other algorithms. While we believe that our trained neural network would exhibit superhuman performance when playing in a series of random games against an intelligent, trained and motivated human, it is not at all clear the performance of two models playing together would be higher than that of two humans doing the same. For once, people might be more likely to coordinate on payoff dominant equilibria. A similar concern would arise in multi-player games where at least two players are not algorithms.


\newpage
\appendix


\section{Space of Games}\label{app:space_of_games}

\begin{figure}[tbh]
	\centering
	\hspace{22pt}
	\begin{subfigure}[b]{0.495\textwidth}
		\centering
		\includegraphics{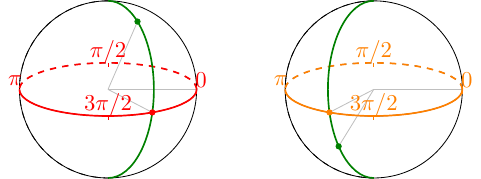}
		\vspace{0.165cm}
		\caption{Space of games, \(\mathcal{U}_{n \times n}\)}
		\label{fig:space_of_games_a}
	\end{subfigure}
	\hspace{-30pt}
	\begin{subfigure}[b]{0.495\textwidth}
		\centering
		\includegraphics{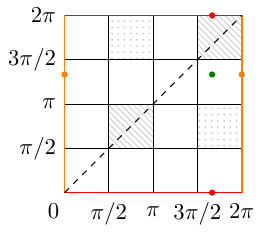}
		\caption{Strategic subspace, \(\mathcal{U}^{\mathrm{N}}_{n \times n}\)}
		\label{fig:space_of_games_b}
	\end{subfigure}
	\caption{A game in \(\mathcal{U}_{2 \times 2}\) can be represented as a pair of points in two spheres (e.g. the pair of green dots in subfigure a). Games in \(\mathcal{U}^{\mathrm{N}}_{2 \times 2}\) are pairs of points along the equators (e.g. the pair of red-orange dots in subfigure a). Pairs of points along the green semicircles are best-reply structure equivalent and are associated to the same element in \(\mathcal{U}^{\mathrm{N}}_{2 \times 2}\), which can be visualized in a torus (subfigure b). }
	\label{fig:space_of_games}
\end{figure}

Let \(\mathbf{1}\) denote the \(n\times1\) all‑ones vector, \(\mathbf{0}\) the \(n\times1\) all‑zeros vector, \(\langle\cdot,\cdot\rangle\) the Frobenius inner product, and \(\|\cdot\|\) the Frobenius norm.  

\subsection*{Space of Games}
We consider the space of \(n \times n\) bimatrix games where each player's payoff matrix is a point in \(\mathbb{R}^{n \times n}\) with norm \(n\) and entries summing to zero. That is, 
\[
	\mathcal{U}_{n \times n} := \mathcal{S}_n \times \mathcal{S}_n,
	\quad \text{where} \quad
	\mathcal{S}_n := \left\{ A \in \mathbb{R}^{n \times n} : \langle \mathbf{1}\mathbf{1}^\top, A \rangle = 0,\ \| A \| = n \right\}.
\]
The space of payoff matrices \(\mathcal{S}_n\) includes exactly one representative for each equivalence class under the equivalence relation on \(\mathbb{R}^{n\times n}\) defined by
\[
	A \sim_{\mathrm{R}} A' \quad\iff\quad \exists\,\alpha>0,\ \beta\in\mathbb{R} \quad\text{s.t.}\quad A' = \alpha A + \beta\,\mathbf{1}\mathbf{1}^\top,
\]
so that any two distinct \(A,A' \in \mathcal{S}_n\) induce different von Neumann–Morgenstern preference orders over lotteries on pure strategy profiles. The only exception is total indifference, as constant payoff matrices \(A\propto \mathbf1\mathbf1^\top\) have no representative in \(\mathcal{S}_n\).

\subsection*{Strategic Subspace}
A relevant measure-zero subspace of \(\mathcal{U}_{n \times n}\) consists of games where each player's payoff matrix is a point in \(\mathbb{R}^{n \times n}\) with norm \(n\) and zero column sums. Formally, 
\[ 
	\mathcal{U}^{\mathrm{N}}_{n \times n} := \mathcal{S}^{\mathrm{N}}_n \times \mathcal{S}^{\mathrm{N}}_n,
	\quad \text{ where } \quad
	\mathcal{S}^{\mathrm{N}}_n := \left\{A \in \mathcal{S}_n : \mathbf{1}^\top A = \mathbf{0}^\top \right\}.
\]
Here \(\mathcal{S}^{\mathrm{N}}_n\) contains exactly one representative for each equivalence class under the equivalence relation on \( \mathcal{S}_n\) defined by
\[
	A \sim_{\mathrm{BR}} A' \quad\iff\quad \exists\,\alpha>0,\ \gamma\in\mathbb{R}^n \quad\text{s.t.}\quad A' = \alpha A + \mathbf{1}\,\gamma^\top,
\]
so that any two distinct \(A,A' \in \mathcal{S}^{\mathrm{N}}_n\) have different best-reply correspondences.
 
\subsection*{Visualizing \(\mathbf{2 \times 2}\) games}

In \(\mathcal{U}_{2\times2}=\mathcal{S}_2\times\mathcal{S}_2 \cong S^2\times S^2\), each payoff matrix corresponds to a unique point on the 2‑sphere \(S^2\).  Hence any game is a pair of points on two spheres embedded in \(\mathbb{R}^3\). 
Restricting to the strategic subspace \(\mathcal{U}^{\mathrm{N}}_{2\times2}=\mathcal{S}^{\mathrm{N}}_2\times\mathcal{S}^{\mathrm{N}}_2\cong S^1\times S^1\), pairs of payoff matrices lie on the equators of the spheres (see \autoref{fig:space_of_games_a}).  As show by \citet{jessie2015},  \(\mathcal{U}^{\mathrm{N}}_{2\times2}\) can be represented by the square \([0,2\pi)\times[0,2\pi)\) with opposite edges identified (see \autoref{fig:space_of_games_b}).


\section{Details of the network architecture.}\label{app:network}

Our network is a multilayer perceptron composed of an input layer that vectorizes the bimatrix game, a sequence of ReLU-activated hidden layers, and a softmax output layer that produces a probability distribution. A detailed description of each layer is given below.

Let \(\mathbf{1}\) and \(\mathbf{0}\) denote the all‑ones vector and the all‑zeros vector of appropriate dimension, respectively.
\subsection*{Input layer:} 
	The input layer receives a bimatrix game \((U^1,U^2)\) and transforms it into a vector in \(\mathbb{R}^{2n^2}\). Formally, 
	\[
	h^{(0)} = \left(\operatorname{vec}(U^1)^\top,\operatorname{vec}(U^2)^\top \right)^\top.
	\]

\subsection*{Hidden layers:} 
	The input layer is followed by \(L\) hidden layers. The first hidden layer takes \(h^{(0)}\) as input and linearly transforms it into a vector of dimension \(d > 2n^2\). A rectified linear unit (ReLU) activation function is then applied. Formally, 
	\[
		{h}^{(1)} = \max\left(\mathbf{0}, \, {W}^{(1)} {h^{(0)}} + {b}^{(1)}\right),
	\]
	where the \(\max\) operator is applied elementwise, and \({W}^{(1)} \in \mathbb{R}^{d \times 2n^2}\), \({b}^{(1)} \in \mathbb{R}^{d}\) are the trainable parameters of the first hidden layer.
	Each of the remaining hidden layers \(l \in \{2, \ldots, L\}\) receives the \(d\)-dimensional output vector of the preceding layer, \({h}^{(l-1)}\), and applies a dimension-preserving linear transformation followed by ReLU activation. Formally,
	\[
		{h}^{(l)} = \max\left(\mathbf{0}, \, {W}^{(l)} {h}^{(l-1)} + {b}^{(l)}\right),
	\]
	where \({W}^{(l)} \in \mathbb{R}^{d \times d}\) and \({b}^{(l)} \in \mathbb{R}^{d}\) are the parameters of the \(l\)-th hidden layer.

\subsection*{Output layer:} 
	The output layer transforms the \(d\)-dimensional output of the last hidden layer, \({h}^{(L)}\), into a point in the \((n-1)\)-dimensional simplex. Formally,
	\[
		{y} = \softmax \left({W}^{(L+1)} {h}^{(L)} + {b}^{(L+1)} \right),
	\]
	where \(\softmax(x)= \exp({x}) / \mathbf{1}^\top \! \exp({x})\) with \(\exp\) operating elementwise, and \({W}^{(L+1)} \in \mathbb{R}^{n \times d}\) and \({b}^{(L+1)} \in \mathbb{R}^{n}\) are the parameters of the output layer.


\section{Behavioural Axioms} \label{app:axioms}

We now define and discuss a number of axioms in turn. In order to do so, we treat the output of our network at convergence as a family of single-valued solution concepts \(f=\{f_n\}_{1<n\leq3}=\{(f_n^1,f_n^2)\}_{1<n\leq3}\), where \(f^i_n : \mathcal{U}_{n \times n} \to \Delta_{n-1}\) for all \(i = 1,2\).

We say that  \(f\) is \textit{symmetric} if it selects the same profile of strategies whenever the role of players is swapped. That is, \(f^1_n(U^1,U^2)=f^2_n(U^2,U^1)\) for all \((U^1,U^2) \in \mathcal{U}_{n \times n}\).  To test this property, we computed the total variation distance between \(f^1_n(U^1,U^2)\) and \(f_n^2(U^2,U^1)\) for all \(2^{17}\) games in the test set with multiple pure equilibria, for \(n=2,3\). We found an average distance (standard deviation) of \(0.006\) \((0.043)\) in \(2\times 2\) games and  \(0.054\) \((0.138)\) in \(3\times 3\) games.  The axiom implies that symmetric equilibria are played in symmetric games (i.e., games where \(U^1=U^2\)). Since the ex-post experience of playing in the two roles is heterogeneous, the symmetry of \(f\) further confirms that the learning is robust in the realisation of games in the training set. 

We say that \(f_n\) satisfies \textit{equivariance} if and only if it selects the same profile of strategies whenever two games differ only because the order of actions has been permuted. That is, if for any pair of permutation matrices \(P, Q \in \mathbb{R}^{n \times n}\) it holds \(f_n^1(P U^1 Q, Q^\top U^2 P^\top) = Pf_n^1(U^1, U^2)\) and \( f_n^2(P U^1 Q, Q^\top U^2 P^\top) = Qf_n^2(U^1, U^2)\). Building on symmetry, we verified that equivariance approximately holds by evaluating for each game in the test set with multiple pure equilibria the output of one of the two networks across the \((n!)^2\) permutations of the game. For each game, we measured the average total variation distance from the centroid strategy across permutations. Then, averaging across games, we obtained an overall average (standard deviation) of \(0.003\) \((0.021)\) in \(2\times 2\) games and  \(0.039\) \((0.077)\) in \(3\times 3\) ones. This finding has a bite in our setting because the order of actions determines the placement of playoffs in the input layer. Therefore, the network could, in principle, coordinate to play a different equilibrium based on the observed order of actions.

We say that \(f_n\) satisfies \textit{invariance to payoff transformations that preserve the best reply structure} if it selects the same profile of strategies following payoff transformations that do not alter the best reply correspondence. That is, if for each player \(i \in \{1,2\}\), whenever \((U^1, U^2), (\tilde{U}^1, \tilde{U}^2) \in \mathcal{U}_{n \times n}\) satisfy \(U^1 \sim_{\mathrm{BR}} \tilde{U}^1\) and \(U^2 \sim_{\mathrm{BR}} \tilde{U}^2\), then \(f_n^i(U^1, U^2) = f_n^i(\tilde{U}^1, \tilde{U}^2)\).\footnote{\(A' \sim_{\mathrm{BR}} A\) if and only if \(A' \in \{\alpha A + \mathbf{1}\gamma^\top \in \mathcal{S}_n : \alpha > 0, \gamma \in \mathbb{R}\}\) (see \autoref{app:space_of_games}).}    
For each game in the test set with multiple pure equilibria we uniformly sampled \(64\) best-reply-equivalent games and computed the average total variation distance of the network outputs with the centroid strategy. Averaging over all games we obtained a mean value (standard deviation) of \(0.026\) \((0.042)\) in \(2\times 2\) games and of \(0.058\) \((0.116)\) in \(3\times 3\) ones.

A solution concept  \(f\) satisfies \textit {monotonicity} if, for all games where it selects a pure equilibrium, the pure equilibrium is still selected if we raise the payoff of players at the equilibrium point. Formally, if \(f_n^1(U^1,U^2), f_n^2(U^1,U^2)\) identifies a pure equilibrium at action profile \(k,z\) and  \(H \in \mathbb{R}^{n \times n}\) has \(H_{k,z} \geq 0\) and all other entries zero, then \(f_n^1(U^1 + H,U^2 + H^\top)\), \(f_n^2(U^1 + H,U^2 + H^\top)\) identifies the same profile. We tested the property by taking all games in the test set with at least a pure equilibrium, generating a pair of random increments uniformly from \([0,1]\), adding them to the payoffs at the equilibrium, and evaluating the difference in behaviour in each such generated game from the centroid computed across all transformed games.  Averaging over all games with multiple pure equilibria we obtained a mean value (standard deviation) of \(0.000\) \((0.001)\) in \(2\times 2\) games and of \(0.000\) \((0.006)\) in \(3\times 3\) ones. In light of the axiomatization by  \citet{Harsanyi1988} (see Theorem 3.9.1), monotonicity and the other three properties above are implied, for \(2\times 2\) games, by the networks selecting the risk-dominant equilibrium in every game.

We say that a solution concept satisfies \textit{independence from strategically irrelevant actions} if it selects the same equilibrium in any two games where one is obtained by adding strictly dominated actions to the other. To formalise this axiom, let \([(U^1,U^2)]_k\) indicate a game restricted to the first \(k\leq n\) actions for both players and \([f_n({U}^1,{U}^2)]_k\) indicate the analogously resticted output of the two networks. Independence from strategically irrelevant actions is satisfied if \(f_n(U^1,U^2)=[f_{n+1}(\tilde{U}^1,\tilde{U}^2)]_n\) whenever \(({U}^1,{U}^2)=[(\tilde{U}^1,\tilde{U}^2)]_n\) and the \((n+1)\)-th actions in \((\tilde{U}^1,\tilde{U}^2)\) for both players are strictly dominated. We tested this axiom by extracting from the test set all \(3\times 3\) games where there was a single strictly dominated action for each player. Relying on symmetry, we then compared the output of one network in each \(3\times 3\) game restricted to the undominated strategies with the output of the same network in the \(2\times 2\) games obtained by eliminating the identified dominated strategies.\footnote{We preferred to use \(3\times 3\) games with dominated strategies as opposed to augmenting \(2 \times 2\) games in order to maintain conditionally uniform sampling of such games in \(\mathcal{U}_{3 \times 3}\). Note that due to the possibility that a small mass is placed on the dominated actions, the resulting restricted strategy is not a distribution. Nonetheless, this has no material effect on computed total variation. } We found an average total variation distance (standard deviation) between \(f^1_2\) and \(f^1_3\) equal to \(0.039\) \((0.130)\) in our sample. This property, which is tested using both the models trained for \(2 \times 2\) and \(3 \times 3\) games, is of independent interest. It establishes coherence between models trained on games of different sizes. The equilibrium selected by the learning process is not affected by the presence of strategically irrelevant actions and would continue to be selected by larger models in games that appear equivalent after dominated strategies are iteratively eliminated. 

A summary of the results for the axioms we tested is presented in \autoref{tab:axioms}.

\begin{table}[tbh]
	\centering
	\footnotesize 
	\hspace{-9pt}
	\begin{tabular}{ccccccc}
    & \multicolumn{6}{c}{\textbf{Axioms}} \\[2pt]
    \cmidrule[1pt](lr){2-7}
    & & \multirow{2}{*}{\textbf{Symmetry}} 
    & \multirow{2}{*}{\textbf{Equivariance}} 
    & \textbf{Best reply} & \multirow{2}{*}{\textbf{Monotonicity}} 
    & \textbf{Independence} \\
    & & & & \textbf{invariance} & & \textbf{irrelev. actions} \\
    \cmidrule[0.6pt](lr){2-7}
    \multirow{6}{*}{\rotatebox{90}{\small \textbf{2x2 Games}}}
    & \textbf{Games} & \small{16564} & \small{16564 $\times$ 4} & \small{16564 $\times$ 64} & \small{16495 $\times$ 64} & \small{8176} \\[4pt]
    & \textbf{Mean Distance} & \small{0.008} & \small{0.005} & \small{0.048} & \small{0.000} & \small{0.043} \\[1pt]
    & \footnotesize{Std. Deviation} & \footnotesize{(0.078)} & \footnotesize{(0.040)} & \footnotesize{(0.066)} & \footnotesize{(0.000)} & \footnotesize{(0.183)} \\[4pt]
    & \textbf{90th Quantile} & \small{0.143} & \small{0.225} & \small{0.314} & \small{0.000} & \small{0.237} \\[4pt]
    & \textbf{99th Quantile} & \small{0.182} & \small{0.285} & \small{0.353} & \small{0.000} & \small{0.998} \\
    \cmidrule[0.6pt](lr){2-7}
    \multirow{6}{*}{\rotatebox{90}{\small \textbf{3x3 Games}}}
    & \textbf{Games} & \small{38594} & \small{38594 $\times$ 36} & \small{38594 $\times$ 64} & \small{30678 $\times$ 64} & --- \\[2pt]
    & \textbf{Mean Distance} & \small{0.071} & \small{0.056} & \small{0.129} & \small{0.001} & --- \\[1pt]
    & \footnotesize{Std. Deviation} & \footnotesize{(0.184)} & \footnotesize{(0.104)} & \footnotesize{(0.108)} & \footnotesize{(0.009)} & --- \\[4pt]
    & \textbf{90th Quantile} & \small{0.838} & \small{0.394} & \small{0.410} & \small{0.014} & --- \\[4pt]
    & \textbf{99th Quantile} & \small{0.929} & \small{0.415} & \small{0.425} & \small{0.018} & --- \\
    \cmidrule[1pt](lr){2-7}
\end{tabular}

	\caption{Statistics for behavioural axioms. The number of games shows for each axiom the number of games in the test set with multiple pure equilibria times (\(\times\)) the number of transformations applied to each game used to test the axiom. Independence of irrelevant actions is tested using both models, so \(3\times 3\) statics are not available. }
	\label{tab:axioms}
\end{table}


\section{Three subspaces of \(\mathcal{U}_{2\times 2}\)}\label{app:subspaces}

\begin{figure}[tbh]
	\centering
	\includegraphics{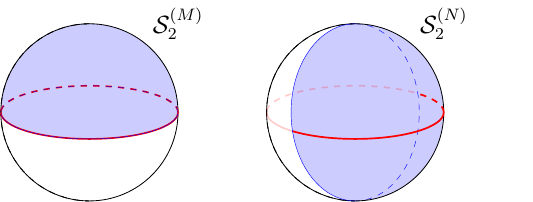}
	\caption{Illustration of the halfspaces \(\mathcal{S}_2^{(M)}\) and \(\mathcal{S}_2^{(N)}\). The set of payoff matrices \(\mathcal{S}_2^{(M)}\) includes all strategic interactions whereas \(\mathcal{S}_2^{(N)}\) excludes half of them.}
	\label{fig:subspaces}
\end{figure}

To characterize different subspaces of \(\mathcal{U}_{n \times n}\), we define for a given matrix \(V \in \mathbb{R}^{n \times n}\) the half-space  
\(
\mathcal{S}_n^{(V)} := \left\{ A \in \mathcal{S}_n : \langle A, V \rangle \geq 0 \right\}.
\)
We then define the subspace associated with a pair of matrices \(V_1,V_2 \in \mathbb{R}^{n \times n}\) as
\[
	\mathcal{U}_{2 \times 2}^{(V_1,V_2)} := \mathcal{S}_2^{(V_1)} \times \mathcal{S}_2^{(V_2)}.
\]
The three subspaces considered are (a) \(\mathcal{U}_{2 \times 2}^{({M,M})}\), (b) \(\mathcal{U}_{2 \times 2}^{({M,N})}\), and (c) \(\mathcal{U}_{2 \times 2}^{({N,N})}\), where  
\[
	M = \begin{pmatrix} -1 & 1 \\ -1 & 1 \end{pmatrix}
	\qquad  \text{and} \qquad
	N = \begin{pmatrix} 1 & -1 \\ -1 & 1 \end{pmatrix}.
\]
Intuitively, \(\mathcal{S}_2^{(M)}\) is the half‐space of \(\mathcal{S}_2\) where payoff matrices have a sum of payoffs in the first column smaller than the sum in the second column.  
Because each equivalence class under \(\sim_{\mathrm{BR}}\) is symmetric about the hyperplane \(\langle M,\,\cdot\rangle=0\), the hemisphere \(\mathcal{S}_2^{(M)}\) contains exactly half of each class (and thus at least one representative of every class). Similarly, \(\mathcal{S}_2^{(N)}\) is the half‐space where the sum along the antidiagonal is smaller than along the diagonal.  
In this case some \(\sim_{\mathrm{BR}}\) classes lie entirely on the negative side of \(\langle B,\,\cdot\rangle=0\), so those classes have no representatives in \(\mathcal{S}_2^{(N)}\). It follows that some strategic interactions (best‐reply correspondences) never occur in \(\mathcal{S}_2^{(N)}\). An illustration is given in \autoref{fig:subspaces}.

\section{Playing larger games} \label{app:larger}   

To confirm that learning takes place in games larger than  \(2\times 2\) and \(3\times 3\), we report our results from training models on \(4\times 4\) and \(5\times 5\) games. The neural networks used are similar to those used in the baseline models, with the exception that each has 2048 neurons per layer. Training took place on \(4,294,967, 296\) games (four times as many) and the batch size used was \(512\) (four times as many).

We evaluated the models using a test set of \(2^{17}\) randomly generated \(4\times 4\) and \(5\times 5\) games. The average maximal normalized regret across players are reported in \autoref{tab:summarylarger}. While these models are less performing than our baseline ones, learning is taking place to a good decree considering the increase in dimensionality. However, it seems natural to expect that, fixing network and training iterations, as the number of actions grows largers the learning problem faced by the networks becomes quickly intractable.

 \begin{table}[h]
	\centering
	\footnotesize 
	\begin{table}[H]
	\centering
	\footnotesize
	\begin{minipage}{.45\linewidth}
		\centering
		\begin{tabular}{cccc}
			\multicolumn{4}{c}{\textbf{$\mathbf{4 \times 4}$ Games}} \\[2pt]
			\toprule
			& \multirow{2}{*}{\textbf{\begin{tabular}[c]{@{}c@{}}All\\[1pt] Games\end{tabular}}} & \multicolumn{2}{c}{\textbf{\# Pure Nash}} \\
			\cmidrule(lr){3-4}
			& & $\mathbf{0}$ & $\mathbf{\geq 1}$ \\
			\midrule
			\textbf{Relative Frequency} & \small 1.00  & \small 0.26 & \small 0.74 \\
			\midrule
			\textbf{Mean MaxReg} & \small 0.02 & \small 0.04 & \small 0.01\\[2pt]
			\textbf{} & \footnotesize (0.04) & \footnotesize (0.04) & \footnotesize (0.03) \\[4pt]
			\midrule
			\textbf{Benchmark} & \small 0.18 & \small 0.18 & \small 0.19 \\[2pt]
			\bottomrule
		\end{tabular}
	\end{minipage}
	\hspace{0.04\linewidth}
	\begin{minipage}{.45\linewidth}
		\centering
		\begin{tabular}{cccc}
			\multicolumn{4}{c}{\textbf{ $\mathbf{5 \times 5}$ Games}} \\[2pt]
			\toprule
			& \multirow{2}{*}{\textbf{\begin{tabular}[c]{@{}c@{}}All\\[1pt] Games\end{tabular}}} & \multicolumn{2}{c}{\textbf{\# Pure Nash}} \\
			\cmidrule(lr){3-4}
			& & $\mathbf{0}$ & $\mathbf{\geq 1}$ \\
			\midrule
			\textbf{Relative Frequency}   & \small 1.00 & \small 0.28 & \small 0.72 \\
			\midrule
			\textbf{Mean MaxReg} & \small 0.03 & \small 0.05 & \small 0.02\\[2pt]
			\textbf{} & \footnotesize (0.05) & \footnotesize (0.05) & \footnotesize (0.05)   \\[4pt]
			\midrule
			\textbf{Benchmark} & \small 0.17 & \small 0.16 & \small 0.17 \\[2pt]
			\bottomrule
		\end{tabular}
	\end{minipage}
	\label{tab:summaryintro}
\end{table}
	\caption{\(4\times 4\) and \(5\times 5\) games. Mean values over the test sets and by number of pure equilibria of the maximum normalized regret across players (MaxReg). Standard deviations in parentheses. Benchmark is the mean maximum normalized regret from random play. Values are rounded.}
	\label{tab:summarylarger}
\end{table}

To investigate whether the output of models trained on larger games agrees with the output of of our baseline models trained on smaller games, we perform a test for invariance to strategically irrelevant actions using the entire family of models (see \autoref{app:axioms} for details). \autoref{tab:comparison} contains the results of the analysis.

\begin{table}[H]
    \centering
    \footnotesize 
    \begin{tabular}{cccc}
        \multicolumn{4}{c}{\textbf{Test of Independence}} \\[2pt]
        \toprule
        & {\(3 \times 3\)} & {\(4 \times 4\)} & {\(5 \times 5\)} \\
        \midrule
        {\(2 \times 2\)} & 0.039 & 0.039 & 0.046 \\[2pt]
        & \footnotesize (0.130) & \footnotesize (0.135) & \footnotesize (0.156) \\[4pt]
        
        {\(3 \times 3\)} &  & 0.095 & 0.094 \\[2pt]
        & \footnotesize  & \footnotesize (0.187) & \footnotesize (0.193) \\[4pt]
        
        {\(4 \times 4\)} &  &  & 0.128 \\[2pt]
        & \footnotesize  & \footnotesize  & \footnotesize (0.201) \\[4pt]
        \bottomrule
    \end{tabular}
    \caption{Test of independence from irrelevant action performed on pairs of models of different dimensions. See \autoref{app:axioms} for details. Standard deviation in parentheses. Numbers are rounded.}
    \label{tab:comparison}
\end{table}

The test scores provide reassurance that larger models play games with \(n\) actions when \(k\) actions for each player are strictly dominated as the smaller models play the same reduced games with \(n-k\) actions.  


\section{Learning Mixed Equilibria} \label{app:mixed}

In this subsection, we report the results of training neural networks assuming the feedback provided is a pure strategy realised from the mixed play of the opponent. We perform this analysis for \(2\times 2\) games under two scenarios. First, we continue to use squared regret as our loss function. Second, we dispense with squared regret and use the absolute value of the regret. Otherwise, the only difference to the baseline model is the initial learning rate which is set to \(\eta_0=0.005\). 

In the first scenario, when regret is squared as in our baseline model, we see that learning appears to converge to a minuimum of the loss. However, as the \autoref{tab:robustness_loss}  below illustrates, the output of the networks is distant from the baseline's one for games with a unique mixed strategy. This is a consequence of the fact that, when agents are not able to observe the entire mixed strategy of the opponent, but only a realisation from it, regret squared is not minimised when agents play a Nash equilibrium in every game. Instead, players will be penalising strategies that produce variance in regret, thus learning a mixed (non-Nash) equilibrium that minimises the squared regret for both players. 

We therefore run the second scenario, where the loss is linear regret. In this case, the loss of both players is minimised if they play a mixed Nash in every game. In this scenario, we observe that the networks fail to converge to a stationary point. MaxReg in the test set seems to settle at 0.18 for games with only a mixed equilibrium, which is worse than the fully random benchmark. Nonetheless, as Table \autoref{tab:robustness_loss}   shows, learning to play pure strategies still takes place.

\begin{table}[h]
	\footnotesize
	\centering
	\begin{table}[H]
	\centering
	\footnotesize
	\begin{minipage}{.45\linewidth}
		\centering
		\begin{tabular}{cccc}
			\multicolumn{4}{c}{\textbf{Squared Regret Loss -- $\mathbf{2 \times 2}$ Games}} \\[2pt]
			\toprule
			& \multirow{2}{*}{\textbf{\begin{tabular}[c]{@{}c@{}}All\\[1pt] Games\end{tabular}}} & \multicolumn{2}{c}{\textbf{\# Pure Nash}} \\
			\cmidrule(lr){3-4}
			& & $\mathbf{0}$ & $\mathbf{\geq 1}$ \\
			\midrule
			\textbf{Relative Frequency} & \small 1.00  & \small 0.13 & \small 0.87 \\
			\midrule
			\textbf{Mean MaxReg} & \small 0.01 & \small 0.06 & \small 0.00\\[2pt]
			\textbf{} & \footnotesize (0.02) & \footnotesize (0.03) & \footnotesize (0.01) \\[4pt]
			\textbf{Mean DistBaseline} & \small 0.07 & \small 0.28 & \small 0.04\\[2pt]
			\textbf{} & \footnotesize (0.14) & \footnotesize (0.14) & \footnotesize (0.12) \\[4pt]
			\bottomrule
		\end{tabular}
	\end{minipage}
	\hspace{0.04\linewidth}
	\begin{minipage}{.45\linewidth}
		\centering
		\begin{tabular}{cccc}
			\multicolumn{4}{c}{\textbf{Linear Regret Loss  -- $\mathbf{2 \times 2}$ Games}} \\[2pt]
			\toprule
			& \multirow{2}{*}{\textbf{\begin{tabular}[c]{@{}c@{}}All\\[1pt] Games\end{tabular}}} & \multicolumn{2}{c}{\textbf{\# Pure Nash}} \\
			\cmidrule(lr){3-4}
			& & $\mathbf{0}$ & $\mathbf{\geq 1}$ \\
			\midrule
			\textbf{Relative Frequency} & \small 1.00  & \small 0.13 & \small 0.87 \\
			\midrule
			\textbf{Mean MaxReg} & \small 0.02 & \small 0.18 & \small 0.00\\[2pt]
			\textbf{} & \footnotesize (0.08) & \footnotesize (0.16) & \footnotesize (0.01) \\[4pt]
			\textbf{Mean DistBaseline} & \small 0.04 & \small 0.27 & \small 0.01\\[2pt]
			\textbf{} & \footnotesize (0.12) & \footnotesize (0.15) & \footnotesize (0.08) \\[4pt]
			\bottomrule
		\end{tabular}
	\end{minipage}
	\label{tab:summaryintro}
\end{table}
	\vspace{-8pt}
	\caption{Squared regret and linear regret minimisation, with realised opponent's action. Mean maximum normalized regret across players (MaxReg) and mean total variation distance to the baseline model (DistBaseline) over \(2^{17}\) games; standard deviation in parentheses. Values are rounded.}
	\label{tab:robustness_loss}
\end{table}

Taken together, these results suggest that some form of regret convexification is essential to learn to play mixed equilibria. It is not the imprecise feedback and the noise due to randomness of the action of the opponent that hampers learning, but the lack of an appropriate loss function. Indeed, assuming a linear loss introduces a decay in performance also if we stick to our baseline scenario with fully observed mixed strategies. We have not performed further analysis, which are the scope of future work, but it is likely that regularising the loss function in the spirit of smooth fictitious play \citep[see][]{Fudenberg1993}, for instance by penalising play of pure strategies, may help.


\section{Robustness to Sampling and Batching}\label{app:sampling}

In a first scenario, we trained the networks on games drawn from a non-uniform distribution on \(2\times 2\). In particular, we sample each payoff matrix from a smooth distribution with full support on \(\mathcal{S}_2\) that assigns higher density to matrices whose four entries are similar in magnitude. The results are in \autoref{tab:robustness_nonuniform} below. We observe that the results are robust to perturbations of the uniform.

\begin{table}[tbh]
	\footnotesize
	\centering
			\begin{tabular}{cccc}
			\multicolumn{4}{c}{\textbf{Non-uniform sampling -- $\mathbf{2 \times 2}$ Games}} \\[2pt]
			\toprule
			& \multirow{2}{*}{\textbf{\begin{tabular}[c]{@{}c@{}}All\\[1pt] Games\end{tabular}}} & \multicolumn{2}{c}{\textbf{\# Pure Nash}} \\
			\cmidrule(lr){3-4}
			& & $\mathbf{0}$ & $\mathbf{\geq 1}$ \\
			\midrule
			\textbf{Relative Frequency} & \small 1.00  & \small 0.13 & \small 0.87 \\
			\midrule
			\textbf{Mean MaxReg} & \small 0.00 & \small 0.01 & \small 0.00\\[2pt]
			\textbf{} & \footnotesize (0.01) & \footnotesize (0.01) & \footnotesize (0.00) \\[4pt]
			\textbf{Mean DistBaseline} & \small 0.01 & \small 0.03 & \small 0.01\\[2pt]
			\textbf{} & \footnotesize (0.05) & \footnotesize (0.04) & \footnotesize (0.06) \\[4pt]
			\bottomrule
		\end{tabular}

	\vspace{-10pt}
	\caption{Non-uniform sampling. Mean maximum normalized regret across players (MaxReg) and mean total variation distance to the baseline model (DistBaseline) over \(2^{17}\) games; standard deviation in parentheses. Values are rounded.}
	\label{tab:robustness_nonuniform}
\end{table}

Next, we look at the effect of batching on learning, focusing on \(2\times 2\) games. We considered two scenarios: online training with no batching and halving the batch size. Except for the number of games played, which is adjusted to keep the same number of optimization steps, and for a lower learning rate of 0.005 chosen for the case of online learning, all other details of the models remain identical to the baseline ones. Results in \autoref{tab:robustness_batching} show that decreasing the batch size has no major effect on learning.

\begin{table}[tbh]
	\footnotesize
	\centering
	\begin{table}[H]
	\centering
	\footnotesize
	\begin{minipage}{.45\linewidth}
		\centering
		\begin{tabular}{cccc}
			\multicolumn{4}{c}{\textbf{Online training -- $\mathbf{2 \times 2}$ Games}} \\[2pt]
			\toprule
			& \multirow{2}{*}{\textbf{\begin{tabular}[c]{@{}c@{}}All\\[1pt] Games\end{tabular}}} & \multicolumn{2}{c}{\textbf{\# Pure Nash}} \\
			\cmidrule(lr){3-4}
			& & $\mathbf{0}$ & $\mathbf{\geq 1}$ \\
			\midrule
			\textbf{Relative Frequency} & \small 1.00  & \small 0.13 & \small 0.87 \\
			\midrule
			\textbf{Mean MaxReg} & \small 0.00 & \small 0.03 & \small 0.00\\[2pt]
			\textbf{} & \footnotesize (0.01) & \footnotesize (0.02) & \footnotesize (0.01) \\[4pt]
			\textbf{Mean DistBaseline} & \small 0.03 & \small 0.08 & \small 0.02\\[2pt]
			\textbf{} & \footnotesize (0.10) & \footnotesize (0.08) & \footnotesize (0.11) \\[4pt]
			\bottomrule
		\end{tabular}
	\end{minipage}
	\hspace{0.04\linewidth}
	\begin{minipage}{.45\linewidth}
		\centering
		\begin{tabular}{cccc}
			\multicolumn{4}{c}{\textbf{Halved batch size -- $\mathbf{2 \times 2}$ Games}} \\[2pt]
			\toprule
			& \multirow{2}{*}{\textbf{\begin{tabular}[c]{@{}c@{}}All\\[1pt] Games\end{tabular}}} & \multicolumn{2}{c}{\textbf{\# Pure Nash}} \\
			\cmidrule(lr){3-4}
			& & $\mathbf{0}$ & $\mathbf{\geq 1}$ \\
			\midrule
			\textbf{Relative Frequency} & \small 1.00  & \small 0.13 & \small 0.87 \\
			\midrule
			\textbf{Mean MaxReg} & \small 0.00 & \small 0.01 & \small 0.00\\[2pt]
			\textbf{} & \footnotesize (0.00) & \footnotesize (0.01) & \footnotesize (0.00) \\[4pt]
			\textbf{Mean DistBaseline} & \small 0.01 & \small 0.04 & \small 0.01\\[2pt]
			\textbf{} & \footnotesize (0.06) & \footnotesize (0.05) & \footnotesize (0.06) \\[4pt]
			\bottomrule
		\end{tabular}
	\end{minipage}
	\label{tab:summaryintro}
\end{table}
	\vspace{-10pt}
	\caption{Different batch sizes. Mean maximum normalized regret across players (MaxReg) and mean total variation distance to the baseline model (DistBaseline) over \(2^{17}\) games; standard deviation in parentheses. Values are rounded.}
	\label{tab:robustness_batching}
\end{table}


\section{Network architecture and Learning Rate} \label{app:architecture}
To verify that our results are robust to changes to the networks’ architecture, we trained three pairs of networks for \(2\times 2\) games with a different architecture.  In one case, we doubled the number of neurons per layer; in the other, we halved them. In a third case, we used asymmetric networks. All other details of the models remained the same. The results in \autoref{tab:robustness_architecture} confirm that even relatively large changes to the network architecture, or asymmetries, do not qualitatively alter the results we obtained in the baseline models.    

\begin{table}[tbh]
	\footnotesize
	\centering
	\begin{tabular}{ccccc}
	&\multicolumn{4}{c}{\textbf{Different architectures -- $\mathbf{2 \times 2}$ Games}} \\[2pt]
	\cmidrule[1pt](lr){2-5}
	&& \multirow{2}{*}{\textbf{\begin{tabular}[c]{@{}c@{}}All\\[1pt] Games\end{tabular}}} & \multicolumn{2}{c}{\textbf{\# Pure Nash}} \\
	\cmidrule(lr){4-5}
	&& & $\mathbf{0}$ & $\mathbf{\geq 1}$ \\
	\cmidrule[0.6pt](lr){2-5}
	&\textbf{Relative Frequency} & \small 1.00  & \small 0.13 & \small 0.87 \\[4pt]
	\cmidrule[0.6pt](lr){2-5}
	\multirow{4}{*}{\small \textbf{\begin{tabular}{c}Halved\\ \# Neurons\end{tabular}}}&\textbf{Mean MaxReg} 
							& \small 0.00 & \small 0.01 & \small 0.00 \\
						   & & \footnotesize (0.00) & \footnotesize (0.01) & \footnotesize (0.00) \\[4pt]
						  &\textbf{Mean DistBaseline} & \small 0.01 & \small 0.04 & \small 0.01 \\
						   & & \footnotesize (0.06) & \footnotesize (0.05) & \footnotesize (0.06) \\[4pt]
	\cmidrule[0.6pt](lr){2-5}
	\multirow{4}{*}{\small \textbf{\begin{tabular}{c}Doubled\\ \# Neurons\end{tabular}}}&\textbf{Mean MaxReg} 
						& \small 0.00 & \small 0.01 & \small 0.00 \\
						   & & \footnotesize (0.01) & \footnotesize (0.01) & \footnotesize (0.00) \\[4pt]
						  &\textbf{Mean DistBaseline} & \small 0.01 & \small 0.03 & \small 0.01 \\
						   & & \footnotesize (0.06) & \footnotesize (0.05) & \footnotesize (0.06) \\[4pt]
	\cmidrule[0.6pt](lr){2-5}
	\multirow{4}{*}{\small \textbf{\begin{tabular}{c}Asymmetric\\Networks\end{tabular}}}&\textbf{Mean MaxReg} 
						& \small 0.00 & \small 0.01 & \small 0.00 \\
						   & & \footnotesize (0.00) & \footnotesize (0.01) & \footnotesize (0.00) \\[4pt]
						  &\textbf{Mean DistBaseline} & \small 0.01 & \small 0.03 & \small 0.01 \\
						   & & \footnotesize (0.05) & \footnotesize (0.04) & \footnotesize (0.05) \\[4pt]
	\cmidrule[1pt](lr){2-5}
\end{tabular}
	\vspace{-8pt}
	\caption{Different networks' architectures. Mean maximum normalized regret across players (MaxReg) and mean total variation distance to the baseline model (DistBaseline) over \(2^{17}\) games; standard deviation in parentheses. Values are rounded.}
	\label{tab:robustness_architecture}
\end{table}

Finally, we consider a scenario with no learning rate decay; that is, \(\alpha=1\).  All other details remained the same as in the baseline scenarios. \autoref{tab:robustness_learning_rate} presents the results. Also in this case, the broad conclusion is that results are robust to changes, even drastic ones, to the learning rate. Note that the networks continue to play close to the baseline scenario.

\begin{table}[tbh]    
	\footnotesize
	\centering
	
		\begin{tabular}{cccc}
			\multicolumn{4}{c}{\textbf{No learning rate decay -- $\mathbf{2 \times 2}$ Games}} \\[2pt]
			\toprule
			& \multirow{2}{*}{\textbf{\begin{tabular}[c]{@{}c@{}}All\\[1pt] Games\end{tabular}}} & \multicolumn{2}{c}{\textbf{\# Pure Nash}} \\
			\cmidrule(lr){3-4}
			& & $\mathbf{0}$ & $\mathbf{\geq 1}$ \\
			\midrule
			\textbf{Relative Frequency} & \small 1.00  & \small 0.13 & \small 0.87 \\
			\midrule
			\textbf{Mean MaxReg} & \small 0.00 & \small 0.02 & \small 0.00\\[2pt]
			\textbf{} & \footnotesize (0.01) & \footnotesize (0.01) & \footnotesize (0.01) \\[4pt]
			\textbf{Mean DistBaseline} & \small 0.02 & \small 0.05 & \small 0.01\\[2pt]
			\textbf{} & \footnotesize (0.08) & \footnotesize (0.06) & \footnotesize (0.08) \\[4pt]			
			\bottomrule
		\end{tabular}

	\vspace{-8pt}
	\caption{No learning rate decay.  Mean maximum normalized regret across players (MaxReg) and mean total variation distance to the baseline model (DistBaseline) over \(2^{17}\) games; standard deviation in parentheses. Values are rounded.}
	\label{tab:robustness_learning_rate}    
\end{table}


\newpage
\bibliographystyle{apalike}
\bibliography{bibfile}

\end{document}